\PassOptionsToPackage{unicode}{hyperref}
\PassOptionsToPackage{hyphens}{url}
\PassOptionsToPackage{dvipsnames,svgnames,x11names}{xcolor}
\documentclass[
]{article}

\usepackage{amsmath,amssymb}
\usepackage{mathrsfs}
\usepackage{iftex}
\ifPDFTeX
  \usepackage[T1]{fontenc}
  \usepackage[utf8]{inputenc}
  \usepackage{textcomp} % provide euro and other symbols
\else % if luatex or xetex
  \usepackage{unicode-math}
  \defaultfontfeatures{Scale=MatchLowercase}
  \defaultfontfeatures[\rmfamily]{Ligatures=TeX,Scale=1}
\fi
\usepackage{lmodern}
\ifPDFTeX\else  
\fi
\IfFileExists{upquote.sty}{\usepackage{upquote}}{}
\IfFileExists{microtype.sty}{% use microtype if available
  \usepackage[]{microtype}
  \UseMicrotypeSet[protrusion]{basicmath} % disable protrusion for tt fonts
}{}
\makeatletter
\@ifundefined{KOMAClassName}{% if non-KOMA class
  \IfFileExists{parskip.sty}{%
    \usepackage{parskip}
  }{% else
    \setlength{\parindent}{0pt}
    \setlength{\parskip}{6pt plus 2pt minus 1pt}}
}{% if KOMA class
  \KOMAoptions{parskip=half}}
\makeatother
\usepackage{xcolor}
\makeatletter
\ifx\paragraph\undefined\else
  \let\oldparagraph\paragraph
  \renewcommand{\paragraph}{
    \@ifstar
      \xxxParagraphStar
      \xxxParagraphNoStar
  }
  \newcommand{\xxxParagraphStar}[1]{\oldparagraph*{#1}\mbox{}}
  \newcommand{\xxxParagraphNoStar}[1]{\oldparagraph{#1}\mbox{}}
\fi
\ifx\subparagraph\undefined\else
  \let\oldsubparagraph\subparagraph
  \renewcommand{\subparagraph}{
    \@ifstar
      \xxxSubParagraphStar
      \xxxSubParagraphNoStar
  }
  \newcommand{\xxxSubParagraphStar}[1]{\oldsubparagraph*{#1}\mbox{}}
  \newcommand{\xxxSubParagraphNoStar}[1]{\oldsubparagraph{#1}\mbox{}}
\fi
\makeatother

\usepackage{longtable,booktabs,array}
\usepackage{calc} % for calculating minipage widths
\usepackage{etoolbox}
\makeatletter
\patchcmd\longtable{\par}{\if@noskipsec\mbox{}\fi\par}{}{}
\makeatother
\IfFileExists{footnotehyper.sty}{\usepackage{footnotehyper}}{\usepackage{footnote}}
\makesavenoteenv{longtable}
\usepackage{graphicx}
\makeatletter
\def\maxwidth{\ifdim\Gin@nat@width>\linewidth\linewidth\else\Gin@nat@width\fi}
\def\maxheight{\ifdim\Gin@nat@height>\textheight\textheight\else\Gin@nat@height\fi}
\makeatother
\setkeys{Gin}{width=\maxwidth,height=\maxheight,keepaspectratio}
\makeatletter
\def\fps@figure{htbp}
\makeatother
\NewDocumentCommand\citeproctext{}{}

\makeatletter
 \let\@cite@ofmt\@firstofone
 \def\@biblabel#1{}
 \def\@cite#1#2{{#1\if@tempswa , #2\fi}}
\makeatother
\newlength{\cslhangindent}
\newlength{\csllabelwidth}
\newenvironment{CSLReferences}[2] % #1 hanging-indent, #2 entry-spacing
 {\begin{list}{}{%
  \setlength{\itemindent}{0pt}
  \setlength{\leftmargin}{0pt}
  \setlength{\parsep}{0pt}
  \ifodd #1
   \setlength{\leftmargin}{\cslhangindent}
   \setlength{\itemindent}{-1\cslhangindent}
  \fi
  \setlength{\itemsep}{#2\baselineskip}}}
 {\end{list}}
\usepackage{calc}

\makeatletter
\@ifpackageloaded{tcolorbox}{}{\usepackage[skins,breakable]{tcolorbox}}
\@ifpackageloaded{fontawesome5}{}{\usepackage{fontawesome5}}
\definecolor{quarto-callout-color}{HTML}{909090}
\definecolor{quarto-callout-note-color}{HTML}{0758E5}
\definecolor{quarto-callout-important-color}{HTML}{CC1914}
\definecolor{quarto-callout-warning-color}{HTML}{EB9113}
\definecolor{quarto-callout-tip-color}{HTML}{00A047}
\definecolor{quarto-callout-caution-color}{HTML}{FC5300}
\definecolor{quarto-callout-color-frame}{HTML}{acacac}
\definecolor{quarto-callout-note-color-frame}{HTML}{4582ec}
\definecolor{quarto-callout-important-color-frame}{HTML}{d9534f}
\definecolor{quarto-callout-warning-color-frame}{HTML}{f0ad4e}
\definecolor{quarto-callout-tip-color-frame}{HTML}{02b875}
\definecolor{quarto-callout-caution-color-frame}{HTML}{fd7e14}
\makeatother
\makeatletter
\@ifpackageloaded{caption}{}{\usepackage{caption}}
\AtBeginDocument{%
\ifdefined\contentsname
  \renewcommand*\contentsname{Table of contents}
\else
  \newcommand\contentsname{Table of contents}
\fi
\ifdefined\listfigurename
  \renewcommand*\listfigurename{List of Figures}
\else
  \newcommand\listfigurename{List of Figures}
\fi
\ifdefined\listtablename
  \renewcommand*\listtablename{List of Tables}
\else
  \newcommand\listtablename{List of Tables}
\fi
\ifdefined\figurename
  \renewcommand*\figurename{Figure}
\else
  \newcommand\figurename{Figure}
\fi
\ifdefined\tablename
  \renewcommand*\tablename{Table}
\else
  \newcommand\tablename{Table}
\fi
}
\@ifpackageloaded{float}{}{\usepackage{float}}
\floatstyle{ruled}
\@ifundefined{c@chapter}{\newfloat{codelisting}{h}{lop}}{\newfloat{codelisting}{h}{lop}[chapter]}
\floatname{codelisting}{Listing}

\usepackage{amsthm}
\theoremstyle{plain}
\newtheorem{proposition}{Proposition}[section]
\theoremstyle{plain}
\newtheorem{theorem}{Theorem}[section]
\theoremstyle{remark}
\AtBeginDocument{}
\newtheorem*{remark}{Remark}

\makeatother
\makeatletter
\@ifpackageloaded{caption}{}{\usepackage{caption}}
\@ifpackageloaded{subcaption}{}{\usepackage{subcaption}}
\makeatother

\ifLuaTeX
  \usepackage{selnolig}  % disable illegal ligatures
\fi
\usepackage{bookmark}

\IfFileExists{xurl.sty}{\usepackage{xurl}}{} % add URL line breaks if available
\hypersetup{
  pdftitle={Inequality Sensitive Optimal Treatment Assignment},
  pdfauthor={Eduardo Zambrano},
  colorlinks=true,
  linkcolor={blue},
  filecolor={Maroon},
  citecolor={Blue},
  urlcolor={Blue},
  pdfcreator={LaTeX via pandoc}}

\title{Inequality Sensitive Optimal Treatment Assignment\footnote{I am
  grateful to Isaiah Andrews for his valuable advice, to Carlos Flores
  for generously sharing his treatment effect bounds Stata files and
  JobCorps data, to Anmol Lakhotia and Michelle Kairy for their
  excellent research assistance, and to audiences at SLU-Umeå, the 2024
  SSCW meetings, and the 2024 European Meetings of the Econometric
  Society for their helpful feedback. This material is based upon work
  supported by the National Science Foundation under Grant \# 2313969.}}
\author{Eduardo Zambrano\footnote{Department of Economics, Cal Poly, San
  Luis Obispo, CA. \url{https://eduardo-zambrano.github.io/}. Email:
  ezambran@calpoly.edu.}}
\date{2025-02-18}

\begin{document}
\maketitle
\begin{abstract}
The egalitarian equivalent, \emph{ee}, of a societal distribution of
outcomes with mean \emph{m} is the outcome level such that the evaluator
is indifferent between the distribution of outcomes and a society in
which everyone obtains an outcome of \emph{ee}. For an inequality averse
evaluator, \(ee<m\). In this paper, I extend the optimal treatment
choice framework in Manski (2024) to the case where the welfare
evaluation is made using egalitarian equivalent measures, and derive
optimal treatment rules for the Bayesian, maximin and minimax regret
inequality averse evaluators. I illustrate how the methodology operates
in the context of the JobCorps education and training program for
disadvantaged youth (Schochet, Burghardt, and McConnell (2008)) and in
Meager (2022)'s Bayesian meta analysis of the microcredit literature.
\end{abstract}

\section{Introduction}\label{introduction}

Starting with the work of Manski (2000), Manski (2004) and following
Wald (1939), Wald (1945), Wald (1971), a large and growing literature in
economics has considered the problem of optimal policy learning and
treatment assignment, rooted in statistical decision theory. For
example, according to Bayesian statistical decision theory, a researcher
represents uncertainty about treatment effects using a prior, identifies
a function that maps incomes to welfare, and selects the policy or
treatment that leads to the highest expected welfare, where the
expectation is taken with respect to that prior. An alternative
decision-theoretic approach avoids the specification of priors and
instead studies rules for learning which policies would be uniformly
satisfactory in terms of welfare (Manski (2004), Manski (2005), Manski
(2007b), Manski (2007a); Schlag (2006); Hirano and Porter (2009); Stoye
(2009), Stoye (2012); Tetenov (2012), Manski and Tetenov (2016), and
Kitagawa and Tetenov (2018)).

The optimal policy learning or treatment assignment problem has also
been considered in parallel literatures developed in both statistics
(Luedtke and Laan (2016), Qian and Murphy (2011), Zhang et al. (2012),
Zhao et al. (2012)), machine learning (Beygelzimer and Langford (2009),
Dudík, Langford, and Li (2011), Li et al. (2012), Jiang and Li (2016),
Thomas and Brunskill (2016), Kallus and Zhou (2018), Strehl et al.
(2010), Swaminathan and Joachims (2015)), and recently at the
intersection of all these fields (Athey and Wager (2021)).

Despite all these developments, the primary approach for policy
evaluation and treatment assignment used both in academia, industry and
policy circles remains estimating average treatment effects of
randomized A/B experiments assessed using the theory of hypothesis
testing. Unfortunately, it is well-known that treatment selection based
on this approach is problematic: First, the approach offers no rationale
for the conventionally used Type I error probabilities (5\% or less) and
Type II error probabilities (10\% to 20\%); second, the approach pays no
attention to the magnitude and distribution of losses to welfare when
errors occur (Manski and Tetenov (2016)). Policymakers selecting from
among different policy options need to account for their distributional
impacts, and existing analyses of randomized policy experiments often
either neglect distributional issues or lack an economic framework for
evaluating them.

The literature on welfare economics is ideally suited for providing the
requisite economic framework, but it has seldom informed the theory and
practice of program evaluation. A notable exception is Kitagawa and
Tetenov (2021), who write: ``{[}The{]} rich and insighful works in
welfare economics have not yet well linked to econometrics and empirical
analysis for policy design.'' In the proposed project, and in line with
the aim of Kitagawa and Tetenov (2021), I establish connections between
the theoretical welfare economics literature and statistical decision
theory.\footnote{An early reference is Dehejia (2005). See also the
  literature reviewed in Section 1.1 of Kitagawa and Tetenov (2021).}
While Kitanawa and Tetenov focus on the class of rank-dependent Social
Preferences (SP) (Blackorby and Donaldson (1978)) and assume that their
welfare criterion is point-identified by the sampling process, here my
focus is specifically on the special case of additively separable SP
(Atkinson (1970)) but consider sampling processes that allows for both
point and partial identification, as in Manski (2024). One should then
view these lines of research as complementary.

The present work is part of a series of papers that seek to develop an
integrated statistical framework for analyzing distributional impacts in
interventions, building on modern welfare economic theory. First,
Fleurbaey and Zambrano (2024) develops a methodology that can help an
evaluator decide among different kinds of SP under certainty, guided by
the tradeoffs between the well-being of individuals the evaluator would
find acceptable. Once the nature of the tradeoffs is precisely
identified, this determines the specific SP one should use. Second,
Zambrano (2024) extends work by Fleurbaey (2010) and identifies the
normative assumptions that an inequality averse social planner, drawing
inferences based on data, would want to maintain in their evaluation of
the treatment effects. The main result in Zambrano (2024) indicates that
these assumptions determine the specific cardinal representation of the
SP under certainty that should be brought into the decision problem
under uncertainty: the egalitarian equivalent (ee) representation, a
concept similar to the certainty equivalent in expected utility theory.
Third, Flores, Kairy, and Zambrano (2024) focuses on developing the
methodology for properly estimating inequality sensitive treatment
effects, and their bounds, together with corresponding uncertainty
estimates for these.

In this paper, I extend the optimal treatment choice apparatus described
in Manski (2024) to the case where the welfare evaluation is made using
the egalitarian equivalent measures described above, and explore the
optimal treatment rules that arise (Section~\ref{sec-point},
Theorem~\ref{thm-AB} and Theorem~\ref{thm-HP}). I also created a
companion website to the paper, available at
\url{https://osf.io/wv5jt/}, which contains interactive visuals, apps
and narratives that explain and motivate our results in simple and
intuitive terms.

\section{Preliminaries}\label{sec-one}

Consider a fixed and finite population of \(n\) individuals, where each
individual \(i\) has a known income \(y_{i} \ge 0\). For ease of
exposition, this Section considers income as an index of individual
advantage, but the results extend immediately to any cardinal,
interpersonally comparable variable. In particular, it is possible to
adjust income for nonmarket aspects of quality of life that individuals
enjoy or endure, and use this adjusted income (usually called
``equivalent income'') as the relevant index instead of ordinary income.

An evaluator has SP that can be represented by a social welfare function
\(W(y_1,...,y_n):=\sum_{i=1}^{n}f\left(y_{i}\right)\) (Atkinson (1970)),
where \(f\) is an increasing function with values taking an interval in
\(\mathbb{R}\cup\left\{ -\infty,+\infty\right\}\). Within this class of
SP, selecting \(f\) pins down the specific SP of the evaluator. This
class of SP is sometimes called generalized-utilitarian, where \(f\)
being and affine function corresponds to the utilitarian SP and \(f\)
being strictly concave corresponds to the prioritarian SP. See, e.g.,
Adler (2022) for discussion.

It is well-known that a given SP has many different cardinal
representations. For instance, for any monotone transformation \(g\),
\(W(y_1,...,y_n)\) and \(g\left(W(y_1,...,y_n)\right)\) generate the
same ranking over income distributions. One such representation is the
\emph{egalitarian equivalent} representation
\(\mathcal{EE}(y_1,...,y_n):=f^{-1}\left(W(y_1,...,y_n)\right)\), which
denotes the level of income \(ee = \mathcal{EE}(y_1,...,y_n)\) such
that, the evaluator would be indifferent between the distribution
\((y_1,...,y_n)\) and \((\underbrace{ee,..., ee}_{n-times}).\) It is
less known, however, that the choice of representation may matter in
practice, that is, in the context of solving an empirical welfare
maximization problem.

To investigate this matter, Zambrano (2024) considers an evaluator that
faces uncertainty about what income distribution arises with a given
treatment. Assume that the evaluator represents the uncertainty through
a finite set of states of the world, \(S = \{s_{1},...,s_{m}\}\). Each
treatment is associated with an income distribution \(y^{s}\) under each
state of the world and a \emph{prospect}, \(y\), collects the income
distributions induced by a given treatment across the \(m\) states of
the world.

The evaluator's problem is then to rank the prospects
\(y = (y_{i}^{s})_{i \in N, s \in S }\), where \(y_{i}^{s}\) describes
the income attained by individual \(i\) in state \(s\),
\(y_{i} = (y_{i}^{s})_{s \in S}\), and
\(y^{s} = (y_{i}^{s})_{i \in N}\). Let
\(\Upsilon \subseteq \mathbb{R}^{nm}\) denote the relevant set of such
prospects over which the evaluation must be made. In this setting, the
SP (a complete, transitive, binary relation) over the set \(\Upsilon\)
is denoted \(R\), with strict preference \(P\) and indifference \(I\).

In what follows it will be useful to refer to the preferences over
prospects the evaluator has for the special case when \(n=1\), and
preferences over the restricted set
\(\Upsilon_{1} \subseteq \mathbb{R}^{m}\) are denoted by \(\succeq\).
One can think of these preferences as those held by an evaluator that is
unconcerned by distributional considerations, and that treats society as
though it consists of a single or representative individual. Therefore,
when comparing two treatments, this evaluator is solely concerned with
the effects this treatment has on this one individual. We take these
preferences as known, perhaps arising from considerations regarding how
the single-person decision problem under uncertainty or ambiguity should
be approached (see, e.g., Stoye (2011) for discussion).

Theorem 3.1 in Zambrano (2024), following Fleurbaey (2010), shows that,
under standard continuity, dominance, and Pareto conditions, the
following is true: if, in the absence of inequality, the evaluator acts
as though there is only one (representative) individual then, in the
presence of both inequality and uncertainty, the evaluator chooses among
treatments by comparing the profiles
\((\mathcal{EE}(y^1),...,\mathcal{EE}(y^m))\) among treatments according
to \(\succeq\).

The interpretation is that the social evaluation can be done as the
single-person evaluation, but one applies the single-person decision
methodology under uncertainty or ambiguity to the \(m-\)dimensional
vector of egalitarian equivalents
\((\mathcal{EE}(y^1),...,\mathcal{EE}(y^m))\). Therefore, given some SP
about known distributions of income in society, one incorporates those
into an optimal treatment assignment framework under uncertainty and
ambiguity, given a set of states of the world, by first aggregating
across individuals for each state, using the egalitarian equivalent
function, and then aggregating across states in whichever way the
evaluator normally does so in single-person problems.

This is significant because it turns out that the choice of
representation of the evaluator's preferences can be consequential in
cases with statistical uncertainty: one would obtain different rankings
over treatments depending on what representation one was using.
Understanding this is crucial before we embark on the development of a
theory of inequality averse optimal treatment assignment, so let's
examine an example in some detail that makes precisely this point.

\subsection{Illustration}\label{illustration}

Consider a situation, reproduced from Zambrano (2024), with three
individuals, two states of the world, and two prospects. Prospect \(a\)
is given by \(\left[\begin{smallmatrix}
  4 & 2\\
  6 & 2\\
  5 & 2
\end{smallmatrix}\right]\) whereas Prospect \(b\) is given by
\(\left[\begin{smallmatrix}
  3 & 1\\
  8 & 1\\
  10 & 1
\end{smallmatrix}\right]\), where rows correspond to individuals and
columns corresponds to states.

\begin{table}

\caption{\label{tbl-panelcalc}Comparing Two Prospects}

\centering{

\subcaption{\label{tbl-a}GM calculations}

\centering{

\begin{tabular}{ccccllccc}
\toprule
 &  & Prospect \(a\) &  &  &  &  & Prospect \(b\) & \\
\midrule
 & State 1 &  & State 2 &  &  & State 1 &  & State 2\\
\(y_{1}\) & 4 &  & 2 &  &  & 3 &  & 1\\
\(y_{2}\) & 6 &  & 2 &  &  & 8 &  & 1\\
\(y_{3}\) & 5 &  & 2 &  &  & 10 &  & 1\\
 & ------ &  & ------ &  &  & ------ &  & ------\\
GM & 4.93 &  & 2 &  &  & 6.21 &  & 1\\
E{[}GM{]} &  & 3.95 &  &  &  &  & 4.48 & \\
\bottomrule
\end{tabular}

}

\subcaption{\label{tbl-b}AM of logs calculations}

\centering{

\begin{tabular}{ccccllccc}
\toprule
 &  & Prospect \(a\) &  &  &  &  & Prospect \(b\) & \\
\midrule
 & State 1 &  & State 2 &  &  & State 1 &  & State 2\\
\(\log{y_{1}}\) & 1.31 &  & 0.69 &  &  & 1.1 &  & 0\\
\(\log{y_{2}}\) & 1.79 &  & 0.69 &  &  & 2.08 &  & 0\\
\(\log{y_{3}}\) & 1.61 &  & 0.69 &  &  & 2.30 &  & 0\\
 & ------ &  & ------ &  &  & ------ &  & ------\\
AML & 1.60 &  & 0.69 &  &  & 1.83 &  & 0\\
E{[}AML{]} &  & 1.29 &  &  &  &  & 1.22 & \\
\bottomrule
\end{tabular}

}

}

\end{table}%

Consider, in addition, an inequality averse evaluator who summarizes the
distribution of income in any given state by the geometric mean (GM).
This means that, given a prospect \(y\), the evaluator computes, for
each state \(s\), the magnitude \((y_1^s y_2^s y_3^s)^{\frac{1}{3}}\).
Assume further that the evaluator is a risk neutral Bayesian decision
maker, with priors of \((\frac{2}{3},\frac{1}{3})\) on states 1 and 2
respectively. Then this evaluator would prefer Prospect \(b\) over
Prospect \(a\) (the expected certainty equivalent for prospect \(a\) is
\(3.95\) whereas for prospect \(b\) is \(4.48\)). Table~\ref{tbl-a}
summarizes these calculations.

However, an equivalent representation of those geometric mean social
preferences in any given state is the arithmetic mean of the logs (AML)
of the incomes. If we were to use this representation, this means that,
given a prospect \(y\), the evaluator computes, for each state \(s\),
the magnitude \({\frac{1}{3}}(\ln y_1^s + \ln y_2^s + \ln y_3^s)\), and
the expected welfare is then \(1.29\) for Prospect \(a\) and \(1.22\)
for Prospect \(b\). The change in representation would thus make us
change how we order the prospects. Table~\ref{tbl-b} summarizes these
calculations.

In a situation like the one described in Table~\ref{tbl-panelcalc}, the
choices made by a minimax regret evaluator also vary depending on
whether one uses the GM or the AML as the representation for the social
preferences, but the choices made by a maximin evaluator will not depend
on the choice of representation. I introduce and discuss these
preferences methods in the context of the welfare evaluation problem
under uncertainty and ambiguity in Section~\ref{sec-man}.

When contemplating which representation to use, it bears noticing that
working with the AML representation of the evaluator's social
preferences under certainty amounts to imputing a degree of risk
aversion to the evaluation that the evaluator does not really have. This
can be seen most easily in the case of a risk neutral Bayesian evaluator
in a single-person evaluation. In this case, using the AML
representation amounts to applying a concave transformation of the data
coming from a single individual before taking expectations, and this
would make the decision maker act as though they are risk
averse.\footnote{With Bernoulli utility function given by the log
  function.} The GM representation, on the other hand, makes no such
imputation.

\section{Egalitarian Equivalent Optimal Statistical
Decisions}\label{sec-man}

Statistical decision theory adds to the above structure by assuming that
the evaluator observes data generated by some sampling distribution and
uses these data to guide their decision. The main difference between
this earlier work and my current analysis is that social preferences in
the present context are represented in accordance with Theorem 3.1 in
Zambrano (2024). This turns out to be of crucial importance under
partial identification and a decision must be made under uncertainty or
ambiguity. I consider here the choices made by an evaluator with
preferences over prospects as in Section~\ref{sec-one} who is
considering two treatments, \(d \in \{a, b\}\) (e.g.~control and
treatment), which respectively lead to the prospects \(y(a)\) and
\(y(b)\) in \(\Upsilon\).

Let \(f\) be a continuous, monotone, strictly concave function. Then the
egalitarian equivalent of the income distribution associated with
\(y^{s}(d)\) is
\(ee^{s}(d):= \mathcal{EE}(y^{s}(d)) = f^{-1}\left(\frac{1}{n} \sum_{i=1}^{n} f(y^{s}_{i}(d))\right)\)
for \(d \in \{a, b\}\).

Assume that the evaluator observes data \(Z^{t}=(Z_{1},\dots,Z_{t})\)
that are independent and identically distributed with \(Z_j \sim P_s\)
on some space \(\mathcal{Z}\). Let \(\mathcal{P}=\{P_s: s\in S\}.\)

As in Section 3 in Manski (2024), I first consider decision making with
the knowledge that econometricians have assumed in identification
analysis. Then the evaluator knows at state \(s\) the probability
distribution \(P_s \in \mathcal{P}\) but not necessarily the underlying
state. In this context, a \emph{treatment rule} is a mapping
\(\delta: \mathcal{P} \rightarrow [0,1]\), which gives the probability
of a (future) individual being assigned to treatment \(b\), given
knowledge of \(P_s\). Then the egalitarian equivalent of treatment rule
\(\delta\) in state \(s\), given \(P_s\), is

\[
ee^{s}(\delta(P_s)) = f^{-1}\left[\delta(P_s)\left(\frac{1}{n}\sum_{i=1}^{n} f\left(y_{is}(b)\right)\right)+(1-\delta(P_s))\left(\frac{1}{n}\sum_{i=1}^{n} f\left(y_{is}(a)\right)\right)\right]
\]

which simplifies to
\(ee^{s}(\delta(P_s)) = f^{-1}\left[\delta(P_s)f(ee^{s}(b))+(1-\delta(P_s))f(ee^{s}(a))\right].\)
For expositional clarity, I label the point-mass assignments
\(ee^{s}(0)\) and \(ee^{s}(1)\) by \(ee^{s}(a)\) and \(ee^{s}(b)\),
respectively.

As a consequence of 3.1 in Zambrano (2024), the treatment assignment
problem boils down to selecting \(\delta(P_s)\) in order to obtain the
most favorable profile \[
(ee^{1}(\delta(P_s)),...,ee^{m}(\delta(P_s)))
\]

according to the preferences \(\succeq\) over the restricted set
\(\Upsilon_{1}\) discussed in Section~\ref{sec-one}, and given \(P_s\).
Let \(\pi\) be a prior probability distribution on \(S\) and, for each
\(P_s \in \mathcal{P}\), let \(S(P_s) \subset S\) denote the truncated
state space obtained with knowledge of \(P_s\). Below I consider three
versions of \(\succeq\), and investigate the characteristics of the
optimal treatment rules according to these criteria.

\begin{equation}\phantomsection\label{eq-bayesian}{
\max_{\delta(P_s) \in [0,1]} E_{\pi}\left[ee^{s}(\delta(P_s)) |S(P_s)\right] 
}\end{equation}

\begin{equation}\phantomsection\label{eq-maximin}{
\max_{\delta(P_s) \in [0,1]} \min_{s \in S(P_s)} ee^{s}(\delta(P_s))  
}\end{equation}

\begin{equation}\phantomsection\label{eq-mmr}{
\min_{\delta(P_s) \in [0,1]} \max_{s \in S(P_s)} \left[ \max \{ee^{s}(a), ee^{s}(b)\} - ee^{s}(\delta(P_s))\right] 
}\end{equation}

Equation~\ref{eq-bayesian} corresponds to the \emph{Bayesian criterion},
Equation~\ref{eq-maximin} corresponds to the \emph{maximin criterion}
and Equation~\ref{eq-mmr} corresponds to the \emph{minimax regret
criterion}, and where the regret of a treatment rule at a state is
defined as the diffference between the most favorable welfare obtainable
given knowledge of the state and the welfare associated with the
treatment rule at that state. Each of these treatment rules encode a
different way of incorporating uncertainty and ambiguity into the
analysis of welfare. A maximin evaluator selects the treatment rule with
the largest worst-case welfare across states; a minimax regret evaluator
anticipates the worst regret of a treatment rule across states, and
focuses on selecting the treatment rule with the smallest worst regret;
and a Bayesian evaluator allows what happens to welfare in the entire
state space to inform the evaluation of a treatment rule, weighted by
the relative likelihood of the different states, as captured by the
evaluator's prior on the state space. For any of these, I assume, for
simplicity, that ties are broken in favor of treatment \(a\).

\subsection{The Role of
Identification}\label{the-role-of-identification}

Two cases of interest arise. In the first case, \(S(P_s) = \{s\}.\) In
this case, we say that the true state is \emph{point identified}. The
true state is \emph{partially identified} if \(S(P_s)\) is a proper
non-singleton subset of \(S\). Define
\begin{equation}\phantomsection\label{eq-eete}{
\tau_{ee}(s):=ee^{s}(b) - ee^{s}(a),
}\end{equation} the \emph{egalitarian equivalent treatment effect
(EETE)} at \(s\).

\subsubsection{Point Identification Results}\label{sec-point}

If the true state is point identified, Equation~\ref{eq-bayesian} and
Equation~\ref{eq-maximin} both reduce to the optimization problem
\(\max_{\delta(P_s) \in [0,1]} ee^{s}(\delta(P_s))\).
Equation~\ref{eq-mmr} reduces to
\(\min_{\delta(P_s) \in [0,1]} [ \max \{ee^{s}(a), ee^{s}(b)\}\)
\(- ee^{s}(\delta(P_s))]\) which is equivalent to
\(\max_{\delta(P_s) \in [0,1]} ee^{s}(\delta(P_s))\). In any of these
three cases, the optimal solution is therefore the same:
\(\delta(P_s) = 1 \left(\tau_{ee}(s) > 0 \right)\), that is, to assign
individuals to treatment \(b\) if \(\tau_{ee}(s) > 0\) and to treatment
\(a\) otherwise. In this case, true egalitarian equivalent welfare at
every state \(s\) is maximized.

\begin{remark}
For fixed \(s\), maximizing \(ee^{s}(\delta(P_s))\) is equivalent to
maximizing

\[\delta(P_s)\left(\frac{1}{n}\sum_{i=1}^{n} f\left(y_{is}(b)\right)\right)+(1-\delta(P_s))\left(\frac{1}{n}\sum_{i=1}^{n} f\left(y_{is}(a)\right)\right).\]

Therefore, when the true state is point identified, the subtlety about
what representation of social preferences one should bring into the
decision analysis does not arise. All representations lead to the same
answer, just as all three decision criteria lead to the same answer.
Both of these conclusions need to be modified when the true state is
partially identified.
\end{remark}

\subsubsection{Partial Identification
Results}\label{partial-identification-results}

If the true state is partially identified, all these criteria generally
yield different answers, and those answers may be sensitive to what
representation of the evaluator's social preferences under certainty one
chooses to adopt. The first point was already made by Manski (2024). The
second point is novel.

Suppose there exist states \(s_{w}, s_{a}\) and \(s_{b}\) such that
\(ee^{s_{w}}(a) = ee^{s_{b}}(a) = \min_{s \in S(P_s)}ee^{s}(a)\),
\(ee^{s_{w}}(b) = ee^{s_{a}}(b) = \min_{s \in S(P_s)}ee^{s}(b)\), and
\(ee^{s_{d}}(d) = \max_{s \in S(P_s)}ee^{s}(d)\) for \(d \in \{a, b\}\).

The interpretation is that state \(s_{w}\) is a worst state for both
treatments, state \(s_a\) is a best state for treatment \(a\) and a
worst state for treatment \(b\), and state \(s_b\) is a best state for
treatment \(b\) and a worst state for treatment \(a\). Assuming the
existence of these states is not needed for the result below, but the
assumption facilitates exposition.

Theorem~\ref{thm-AB} below extends the results in Sections 5-1-5.3 of
Manski (2024) to the present setting.

\begin{theorem}[]\protect\hypertarget{thm-AB}{}\label{thm-AB}

Assume that the true state \(s\) is partially identified, the set
\(\{(ee^{s}(a),ee^{s}(b))\}_{s \in S(P_s)}\) is bounded, and that
\(s_{w},s_{a},s_{b} \in S(P_s)\). Then

\begin{itemize}
\item
  The solution to the Bayesian decision problem is
  \[\delta^{B}(P_s) = 1 \left(E_{\pi}\left[\tau_{ee}(s)|S(P_s)\right] > 0\right).\]
\item
  The solution to the maximin decision problem is
  \[\delta^{M}(P_s) = 1 \left(\tau_{ee}(s_w) > 0 \right).\]
\item
  The solution to the minimax regret decision problem is
  \(\delta^{R}(P_s) \in (0,1)\) such that
  \[ee^{s_{a}}(a) - ee^{s_{a}}(\delta^{R}(P_s)) = ee^{s_{b}}(b) - ee^{s_{b}}(\delta^{R}(P_s)).\]
\end{itemize}

\end{theorem}

All proofs are in the Appendix.

\subsection{Application: The Choice Between a Status Quo Treatment and
an Innovation When Outcomes Are Binary under Partial
Identification}\label{sec-innovation}

In this Section, I aim to illustrate how the optimal treatment rules
used by inequality averse evaluators compare to those used by their
inequality neutral counterparts in a concrete setting under partial
identification. My starting point below is a variation on the binary
outcome example in Manski (2004), p.~1226, and Manski (2019), p.~301,
also studied in Stoye (2009), p.~72.

An evaluator with preferences over prospects in \(\Upsilon\) as in
Section~\ref{sec-one} is considering two treatments, \(d \in \{a, b\}\),
with \(y_{i}^{s}(d) \in \{\underline{p},\bar{p}\}\) for \(d = a, b\) and
\(\underline{p},\bar{p} \in (0,1]\). The evaluator knows the outcome
distribution of the status quo treatment, \(y(a)\), but does not know
the outcome distribution of the innovation, \(y(b)\). Let
\(P(y^{s}) = \frac{1}{n} \sum_{i=1}^{n} 1_{\{y_{i}^{s}=\bar{p}\}}\),
\(p(a) = P(y(a))\) and \(p^{s}(b) = P(y^{s}(b))\) for \(s = 1,...,m\).
Further, assume that \(s_{a}, s_{b} \in S\), where
\(0 = P(y^{s_{a}}(b)) < p(a) < P(y^{s_{b}}(b)) = 1\).

Let
\(\mathcal{EE}(x_{1},...,x_{n}) = f^{-1}(\frac{1}{n} \sum_{i=1}^{n} f(x))\),
with \(f\) continuous, strictly increasing, strictly concave, with
\(f(\underline{p}) = 0\) and \(f(\bar{p}) = 1\). Then the egalitarian
equivalent of the outcome distribution associated with \(y^{s}\) is
\(\mathcal{EE}(y^{s}) = f^{-1}(P(y^{s})f(\bar{p}) + (1 - P(y^{s}))f(\underline{p})) = f^{-1}(P(y^{s})).\)

In this context, the interpretation is that
\(\mathcal{\mathcal{EE}}(y^{s})\) is the notional level of the outcome
variable such that, if everyone in society had that outcome level, then
the evaluator would be indifferent between that notional society and the
actual society, where \(P(y^{s})\) of the individuals have an outcome of
\(\bar{p}\) and \(1 - P(y^{s})\) of the individuals have an outcome of
\(\underline{p}\).

Let \(\delta\) be the fraction of the population to be assigned to the
treatment. Then the egalitarian equivalent of treatment rule \(\delta\)
in state \(s\) is
\(ee^{s}(\delta) = f^{-1}(p(a) + (p^{s}(b) - p(a))\delta)\), where I
simply write \(ee(a)\) for \(ee^{s}(a).\)

As in Section~\ref{sec-man} above, the evaluation boils down to the
statistical comparison of the profiles
\((ee^{1}(\delta),...,ee^{m}(\delta))\) as one varies \(\delta\) from
zero to one. I now show how the three statistical decision theories
discussed above would approach this evaluation.

\subsubsection{The Bayesian Evaluators}\label{the-bayesian-evaluators}

An inequality averse (\(IA\)) Bayesian evaluator would choose \(\delta\)
to maximize \(\sum_{s=1}^{m} ee^{s}(\delta)\pi(s)\) for some prior
\(\pi\) on \(S\). For reference, in this context an inequality neutral
(\(IN\)) Bayesian evaluator chooses \(\delta\) to maximize
\(\sum_{s=1}^{m} (p(a) + (p^{s}(b) - p(a))\delta) \pi(s)\). The solution
to the \(IN\) Bayesian evaluator's problem is: \(\delta^*_{in} = 1\) if
\(E_{\pi}[p(b)] := \sum_{s=1}^{m} p^{s}(b) \pi(s) > p(a)\) and
\(\delta^*_{in} = 0\) if \(E_{\pi}[p(b)] \le p(a)\) (Manski (2004),
p.~1228). The following is true:

\begin{proposition}[]\protect\hypertarget{prp-IN}{}\label{prp-IN}

If the \(IN\) Bayesian evaluator accepts the innovation, so will the
\(IA\) Bayesian evaluator. However, the converse need not hold.
Furthermore, if either Bayesian evaluator accepts the innovation, they
set \(\delta = 1\), and if they don't accept it, they set
\(\delta = 0\).

\end{proposition}

\begin{figure}

\centering{

\includegraphics[width=0.7\textwidth,height=\textheight]{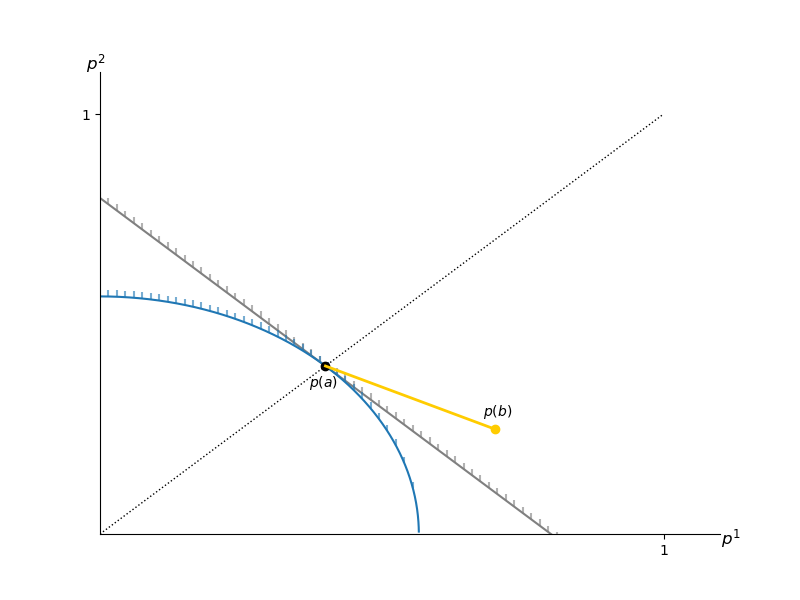}

\includegraphics[width=0.7\textwidth,height=\textheight]{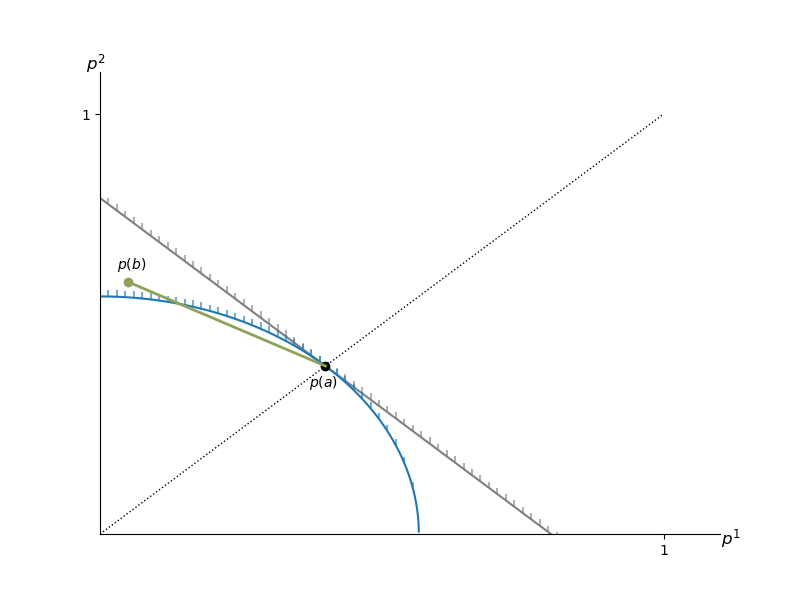}

\includegraphics[width=0.7\textwidth,height=\textheight]{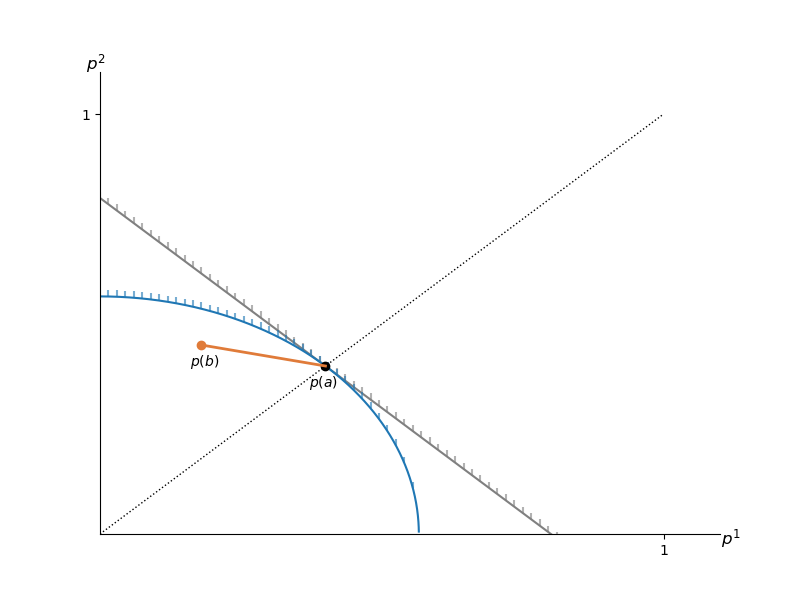}

}

\caption{\label{fig-Baye}The Bayesian evaluators}

\end{figure}

Figure~\ref{fig-Baye} provides the intuition for the \(m = 2\) case,
given a prior \(\pi\) over the two states. The axes measure the
proportions \(P(y)\) in both states of the world for any prospect \(y\).
The gray line denotes all prospects \(y\) with \(E_{\pi}[P(y)] = p(a).\)
Therefore, the \(IN\) Bayesian evaluator accepts any prospect in the
shaded gray area. The blue line denotes all prospects \(y\) with
\(E_{\pi}[\mathcal{EE}(y)] = ee(a)\). Therefore, the \(IA\) Bayesian
evaluator accepts any prospect in the shaded blue area. The segment
connecting \(p(a)\) (corresponding to \(\delta = 0\)) to \(p(b)\)
(corresponding to \(\delta = 1\)) contains all the treatment rules the
evaluator chooses from. Thus, in the top panel all the feasible
treatment rules are in the yellow segment, and both Bayesian evaluators
set \(\delta^*_{in} = 1\). In the middle panel, all the feasible
treatment rules are in the green segment, the \(IN\) Bayesian evaluator
rejects the treatment, setting \(\delta^*_{in} = 0\), but the \(IA\)
Bayesian evaluator accepts the treatment, setting \(\delta^*_{ia} = 1\).
In the bottom panel, all the feasible tretament rules are in the orange
segment, and both evaluators reject the treatment, setting
\(\delta^*_{in} = \delta^*_{ia} = 0\).

\subsubsection{The Minimax Regret Evaluators}\label{sec-innovation-MMR}

A minimax regret evaluator would choose \(\delta\) to minimize
\(R(\delta):= max_{s \in S} R(\delta,s)\), where the \(IA\) evaluator
defines \(R(\delta,s)\) by
\(R_{ia}(\delta,s) := \max \{ee(a), ee^{s}(b)\} - ee^{s}(\delta)\)
whereas \(R(\delta,s)\) for the \(IN\) evaluator simplifies to
\(R_{in}(\delta,s) := (1 - \delta)[p^{s}(b) - p(a)] 1_{\{p^{s}(b) \ge p(a)\}} + \delta[p(a) - p^{s}(b)] 1_{\{p^{s}(b) \le p(a)\}}\).
The solution to the \(IN\) minimax regret problem is:
\(\delta^{*}_{in} = 1 - p(a)\). (Stoye (2009), p.~73).

\begin{proposition}[]\protect\hypertarget{prp-IA}{}\label{prp-IA}

The \(IA\) minimax regret evaluator chooses to treat a larger fraction
of the population with the innovation than the \(IN\) minimax regret
evaluator.

\end{proposition}

\begin{figure}

\begin{minipage}{0.50\linewidth}
\includegraphics{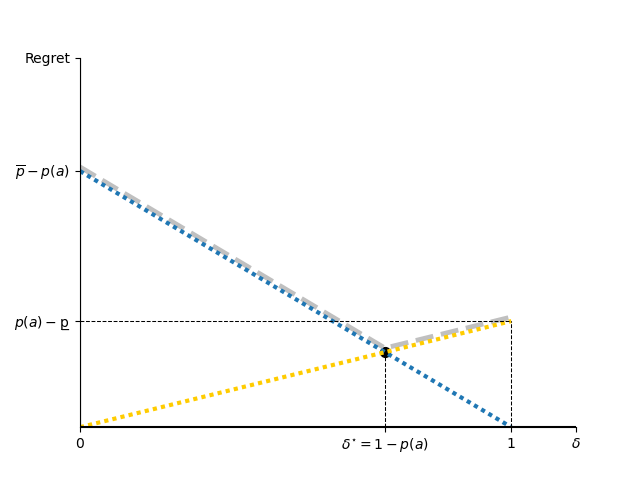}\end{minipage}%
\begin{minipage}{0.50\linewidth}
\includegraphics{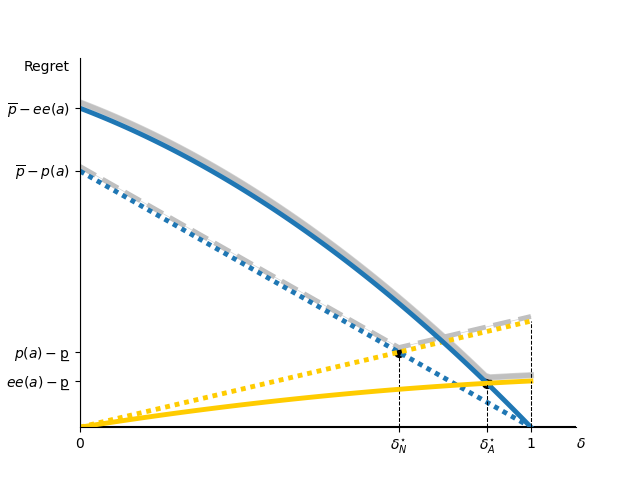}\end{minipage}%

\caption{\label{fig-two}The minimax regret evaluators}

\end{figure}

Figure~\ref{fig-two} provides the intuition for the result. In the left
panel, the dashed yellow line corresponds to \(R_{in}(\delta,s_{a})\),
the dashed blue line corresponds to \(R_{in}(\delta,s_{b})\), the dashed
gray line corresponds to \(R_{in}(\delta)\). This function is minimized
at \(\delta^{*}_{in} = 1 - p(a)\). In the right panel, the solid yellow
line corresponds to \(R_{ia}(\delta,s_{a})\), the solid blue line
corresponds to \(R_{ia}(\delta,s_{b})\), the solid gray line corresponds
to \(R_{ia}(\delta)\). This function is minimized at
\(\delta^{*}_{ia} > 1 - ee(a)\). Since \(ee(a) < p(a)\), the result
follows.

\begin{remark}
Regret here is non-linear in \(\delta\), as in Kitagawa, Lee, and Qiu
(2024). The difference between their setting and the present setting is
that, in their setting, they apply a non-linear transformation to an
otherwise standard measure of inequality neutral regret, whereas here
the non-linearity stems directly from the attitudes towards inequality
of the evaluator. I view these lines of work as complementary, and
looking further into their similarities and differences is left for
future work.
\end{remark}

\subsubsection{The Maximin Evaluators}\label{the-maximin-evaluators}

An \(IA\) maximin evaluator chooses \(\delta\) to maximize
\(min_{s \in S} \{ee^{s}(\delta)\}\), whereas the \(IN\) maximin
evaluator maximizes instead
\(min_{s \in S} \{(p(a) + (p^{s}(b) - p(a))\delta)\}\). The solution for
the \(IN\) maximin evaluator is \(\delta^{*} = 0\) (Manski (2004),
p.~1228) but, because \(f^{-1}\) is monotone, both problems share the
same solution.

\begin{remark}
In any of the three cases considered above (Bayesian, minimax regret, or
maximin), one obtains the same solution when ranking prospects by
applying the decision criteria to either
\((y^{s}_{i})_{i \in N, s \in S}\) or
\((f(y^{s}_{i}))_{i \in N, s \in S}\) whereas this is not so when one
applies the decision criteria to \((\mathcal{EE}(y^{s}))_{s \in S}\).
This is not only a reminder that the choice of representation of the
social preferences under certainty matters (the main point of the
illustration in Section~\ref{sec-one}), but it also highlights that
certain representations will make inequality aversion play no role in
the analysis, even as we intend for it to do so.
\end{remark}

\subsection{Application: Minimizing Egalitarian Equivalent Regret at
JobCorps}\label{sec-jobcorps}

JobCorps (Schochet, Burghardt, and McConnell (2008)) is a widely studied
education and training program for disadvantaged youth. Despite it being
a randomized intervention, estimating the effect of the program on
applicant's wages is difficult due to the fact that the evaluator only
observes wages for those individuals who are employed. The implication
of this, in the context of the present paper, is that, to the extent
that the outcome variable of interest for the evaluator is wages, one
may not be able to point identify at state \(s\) the objects
\(ee^{s}(a)\) and \(ee^{s}(b)\) and therefore \(\tau_{ee}(s)\) is also
not point identified at \(s\). However, partial identification may be
achievable, under relatively mild assumptions.

Flores, Kairy, and Zambrano (2024) adapts the bounds analysis of
Horowitz and Manski (2000), Lee (2009), and Chen and Flores (2015) in
order to arrive at a relevant set of lower and upper bounds for
\(ee^{s}(a)\) and \(ee^{s}(b)\). While the goal in Flores, Kairy, and
Zambrano (2024) is to use those magnitudes to obtain bounds on the
\(EETE\), below I use them to determine the optimal treatment assignment
according to the minimax regret criterion when the evaluator is
inequality averse. I abstract from estimation problems in this
sub-section and treat these bounds as recoverable in a large sample and
therefore known.\footnote{The proper estimation metodology of these
  effects, together with their bounds and uncertainty estimates is the
  focus of Flores, Kairy, and Zambrano (2024).}

In order to arrive at inequality sensitive estimations of the effect of
the treatment, one needs to specify the function \(f\) that captures the
attitudes towards inequality of the evaluator. In this application, I
consider an evaluator with \(f(y) = \frac{y^{1-\gamma}}{1-\gamma}\),
where \(\gamma \ge 0\) is an inequality aversion parameter. I allow for
two kinds of evaluators: \(\gamma = 0\) and \(\gamma = 2\). To interpret
these choices, consider that, when \(\gamma = 2\), the evaluator would
wish, in a two-person evaluation, to protect at least \(50\%\) of an
individual's wage regardless of what happens to the wages of the other
individual. For reference, the egalitarian equivalent measure that
corresponds to \(\gamma = 2\) is the harmonic mean, and
\(\gamma \rightarrow 1\) and \(\gamma =0\) correspond, respectively, to
the egalitarian equivalent measures given by the geometric and
arithmetic means. Either of those last two choices corresponds to a
level of protected wages equal to zero. Loosely, values of \(\gamma\)
greater than one essentially protect those with the lowest wages from
losing everything should the better off in terms of wages gain
disproportionately from an intervention. See Fleurbaey and Zambrano
(2024) for details.

\begin{table}

\caption{\label{tbl-panelcalc-ota}Egalitarian Equivalent Bounds}

\centering{

\subcaption{\label{tbl-osd_a}Inequality neutral evaluator:
\(\gamma = 0\)}

\centering{

\begin{tabular}{ccccc}
\toprule
 & \(ee^L(a)\) & \(ee^U(a)\) & \(ee^L(b)\) & \(ee^U(b)\)\\
\midrule
Horowitz and Manski & 5.4 & 10.7 & 6.1 & 12\\
Lee & 7.9 & 7.9 & 7.5 & 8.7\\
Chen and Flores & 7.9 & 7.9 & 8.3 & 8.7\\
\bottomrule
\end{tabular}

}

\subcaption{\label{tbl-osd-b}Inequality averse evaluator:
\(\gamma = 2\)}

\centering{

\begin{tabular}{ccccc}
\toprule
 & \(ee^L(a)\) & \(ee^U(a)\) & \(ee^L(b)\) & \(ee^U(b)\)\\
\midrule
Horowitz and Manski & 4.1 & 9 & 3.8 & 8.6\\
Lee & 6.6 & 6.6 & 6.5 & 7.7\\
Chen and Flores & 6.6 & 6.6 & 6.8 & 7.7\\
\bottomrule
\end{tabular}

}

}

\end{table}%

\begin{figure}

\centering{

\includegraphics{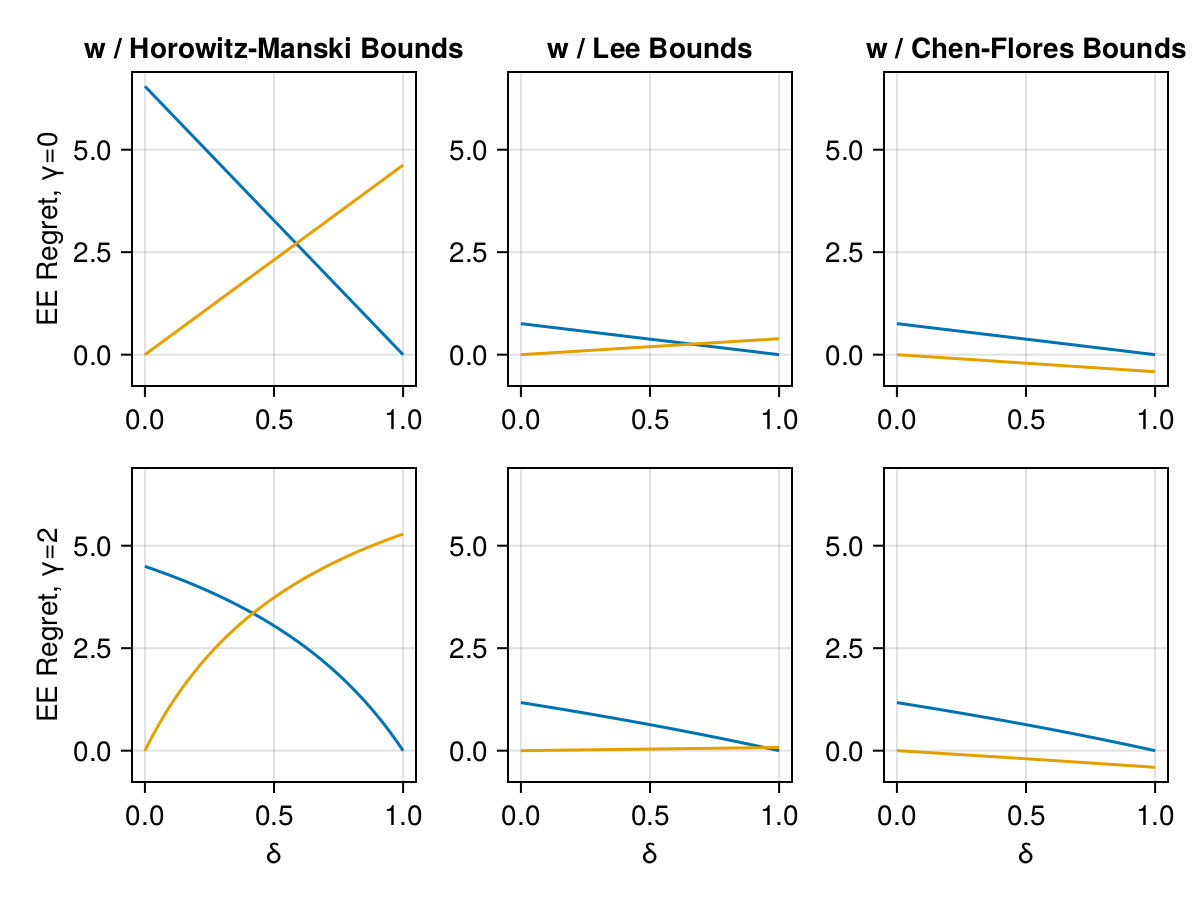}

}

\caption{\label{fig-ota}Minimizing Egalitarian Equivalent Regret at
JobCorps}

\end{figure}%

Table~\ref{tbl-panelcalc-ota} reports Horowitz and Manski, Lee, and Chen
and Flores bounds for \(ee^{s}(a)\) and \(ee^{s}(b)\), for
\(\gamma = 0\) and \(\gamma = 2\), when the outcome variable is hourly
wages at week 208 after random assignment. Those bounds can then be used
to obtain the corresponding optimal treatment assignments, using
Theorem~\ref{thm-AB}. Figure~\ref{fig-ota} describes how these
assignments are obtained. In each graph, the blue line denotes
\(R(\delta, s_b):=ee^{s_b}(b) - ee^{s_b}(\delta)\), with
\(R(0, s_b) = ee^U(b) - ee^L(a)\) (computed using the estimates in
Table~\ref{tbl-panelcalc-ota}), and \(R(1, s_b)=0\). In turn, the yellow
line denotes \(R(\delta, s_a):=ee^{s_a}(a) - ee^{s_a}(\delta)\) with
\(R(0, s_a)=0\) and \(R(1, s_a) = ee^U(a) - ee^L(b)\) (also computed
using the estimates in Table~\ref{tbl-panelcalc-ota}). The intersecion
of these lines in the four leftmost graphs in Figure~\ref{fig-ota}
corresponds to the treatment rule that minimizes worst regret. On the
two graphs on the right in Figure~\ref{fig-ota}, regret is always worst
along \(R(\delta, s_b)\), as the blue line is uniformly above the yellow
line, and in those two cases worst regret is minimized at
\(\delta = 1\). Table~\ref{tbl-OTA_values} reports the resulting optimal
treatment assignments.

\begin{longtable}[]{@{}
  >{\centering\arraybackslash}p{(\columnwidth - 6\tabcolsep) * \real{0.2500}}
  >{\centering\arraybackslash}p{(\columnwidth - 6\tabcolsep) * \real{0.2500}}
  >{\centering\arraybackslash}p{(\columnwidth - 6\tabcolsep) * \real{0.2500}}
  >{\centering\arraybackslash}p{(\columnwidth - 6\tabcolsep) * \real{0.2500}}@{}}
\caption{Optimal Treatment Assignment, \(\delta\), under Partial
Identification}\label{tbl-OTA_values}\tabularnewline
\toprule\noalign{}
\begin{minipage}[b]{\linewidth}\centering
\end{minipage} & \begin{minipage}[b]{\linewidth}\centering
Horowitz and Manski
\end{minipage} & \begin{minipage}[b]{\linewidth}\centering
Lee
\end{minipage} & \begin{minipage}[b]{\linewidth}\centering
Chen and Flores
\end{minipage} \\
\midrule\noalign{}
\endfirsthead
\toprule\noalign{}
\begin{minipage}[b]{\linewidth}\centering
\end{minipage} & \begin{minipage}[b]{\linewidth}\centering
Horowitz and Manski
\end{minipage} & \begin{minipage}[b]{\linewidth}\centering
Lee
\end{minipage} & \begin{minipage}[b]{\linewidth}\centering
Chen and Flores
\end{minipage} \\
\midrule\noalign{}
\endhead
\bottomrule\noalign{}
\endlastfoot
\(\gamma = 0\) & 0.59 & 0.66 & 1 \\
\(\gamma = 2\) & 0.42 & 0.95 & 1 \\
\end{longtable}

To better understand the results from Table~\ref{tbl-OTA_values} note
that, as we move, for each row, from the leftmost to the rightmost
column in the table, the fraction of individuals assigned to treatment
\(b\) grows, until the fraction reaches 1. The reason why this happens
is that the move from the leftmost to the rightmost column in
Table~\ref{tbl-OTA_values} corresponds to a progressive reduction in the
size of the identified set, until it no longer contains zero (Chen and
Flores (2015)).

From Table~\ref{tbl-OTA_values} we also learn that, unlike in the
application discussed in Section~\ref{sec-innovation}, it is not always
the case that the inequality averse evaluator assigns a comparative
larger fraction of the population to treatment \(b\), relative to the
inequality neutral evaluator. The differences between these applications
stem from the following: In the application from
Section~\ref{sec-innovation}, the introduction of inequality aversion
increases the worst regret of treatment \(a\) but decreases the worst
regret of treatment \(b\), as shown on the right panel in
Figure~\ref{fig-two}. This has the unambiguous effect, as discussed in
Section~\ref{sec-innovation-MMR}, of increasing the fraction assigned to
treatment \(b\) as inequality aversion grows. In the present
application, however, inequality aversion may increase or decrease the
worst regret of either treatment. This is so because worst regret is
computed as the difference of two egalitarian equivalent measures, each
of which shrink with inequality aversion. Which of the two measures
shrinks faster then determines whether worst regret grows or shrinks for
a given treatment option as inequality aversion increases.

Let's take a look at how this plays out in practice in the context of
the JobCorps application. First, consider the case of the Horowitz and
Manski bounds. Given those bounds, the worst regret from treatment \(a\)
shrinks from \(\$12-\$5.4=\$6.6\) per hour to \(\$8.6-\$4.1=\$4.5\) per
hour as we move from \(\gamma=0\) to \(\gamma=2\)
(Table~\ref{tbl-panelcalc-ota}). This is driven by the fact that
\(ee^U(b)\) drops by more (\(\$12-\$8.6=\$3.4\)) than what \(ee^L(a)\)
drops by (\(\$5.4-\$4.1=\$1.3\)) as we move from \(\gamma=0\) to
\(\gamma=2\). On the other hand, the worst regret from treatment \(b\)
grows from \(\$10.7-\$6.1=\$4.6\) per hour to \(\$9-\$3.8=\$5.2\) per
hour as we move from \(\gamma=0\) to \(\gamma=2\). This is driven by the
fact that \(ee^U(a)\) drops by less (\(\$10.7-\$9=\$1.7\)) than what
\(ee^L(b)\) drops by (\(\$6.1-\$3.8=\$2.3\)) as we move from
\(\gamma=0\) to \(\gamma=2\). The drop in worst regret for treatment
\(a\), coupled with the rise in worst regret for treatment \(b\), as we
move from \(\gamma=0\) to \(\gamma=2\), therefore causes the optimal
treatment assignment to move from \(\delta=0.59\) to \(\delta = 0.42\)
(Table~\ref{tbl-OTA_values}).

The case with Lee bounds illustrates the opposite situation, where the
worst regret from treatment \(a\) grows from \(\$8.7-\$7.9=\$0.8\) per
hour to \(\$7.7-\$6.6=\$1.1\) per hour as we move from \(\gamma=0\) to
\(\gamma=2\) (Table~\ref{tbl-panelcalc-ota}). On the other hand, the
worst regret from treatment \(b\) shrinks from \(\$7.9-\$7.5=\$0.4\) per
hour to \(\$6.6-\$6.5=\$0.1\) per hour as we move from \(\gamma=0\) to
\(\gamma=2\). The worst regret for treatment \(b\) is much lower than
the worst regret for treatment \(a\) for \(\gamma=0\), which is why the
optimal treatment assignment in this case is above \(0.5\), at
\(\delta=0.66\) (Table~\ref{tbl-OTA_values}). The rise in worst regret
for treatment \(a\) coupled with the drop in worst regret for treatment
\(b\), as we move from \(\gamma=0\) to \(\gamma=2\), causes the optimal
treatment assignment to move further up, from \(\delta=0.66\) to
\(\delta = 0.95\).

Finally, the case with the Chen and Flores bounds nicely illustrates
that both the inequality neutral and inequality averse evaluators may
agree in many cases of interest. In particular, whenever both the upper
and lower bounds of the egalitarian equivalent treatment effect share
the same sign, both evaluators will assign the same treatment: treatment
\(a\) if they share a negative sign, and treatment \(b\) if they share a
positive sign. Given the treatment effect bounds implied by
Table~\ref{tbl-panelcalc-ota}, both evaluators will assign everyone to
treatment \(b\) in this case (Table~\ref{tbl-OTA_values}).

\section{Finite Sample Analysis}\label{finite-sample-analysis}

In this Section, I consider the situation where the evaluator does not
know the sampling distribution but observes finite data,
\(Z^t \in \mathcal{Z^t}\), that are informative about \(s\). A
\emph{statistical treatment rule}
\(\delta: \mathcal{Z^t} \rightarrow [0,1]\), gives the probability of a
(future) individual being assigned to treatment \(b\), given \(Z^t\). To
the extent that knowledge of finite data does not shrink the state
space, the only decision problem worth revisiting is that of the
Bayesian evaluator, which becomes

\begin{equation}\phantomsection\label{eq-bayesianpost}{
\max_{\delta(Z^t) \in [0,1]} E_{\pi}\left[ee^{s}(\delta(Z^t)) |Z^t\right] 
}\end{equation}

and has as solution
\(\delta^{B}(Z^t) = 1 \left(E_{\pi}\left[\tau_{ee}(s)|Z^t\right] > 0\right)\),
a solution analogous to that obtained in Theorem~\ref{thm-AB} for the
Bayesian evaluator.

\subsection{Application: A Bayesian Meta Analysis of the Microcredit
Literature}\label{sec-meager}

To illustrate how the solution of the problem described by
Equation~\ref{eq-bayesianpost} can be used in practice, I now turn to an
application where the evaluator is Bayesian and has access to the
outcome of several related randomized experiments. In particular, I
examine Meager (2022), who estimates posterior distributions of the
effect of microcredit interventions on profit, consumption and other
variables using data from randomized trials that expand access to
microcredit in seven countries. I take the distribution of consumption
before and after treatment to be the primary object of analysis and for
this reason I focus below on the five countries in Meager's meta study
for which consumption data is available.

Meager reports considerable treatment effect heterogeneity, with large
segments of the distribution of consumption nearly unaffected by the
policy (from the 5-th to the 75-th percentiles), together with large yet
uncertain differences on the upper tails of the distribution of
consumption of the treatment and control groups, especially within the
group of households with previous business experience. Meager states
that, given that the treatment will probably increase inequality, ``the
social welfare effects of microcredit are likely to be complex.''
(Meager (2022), p.~1821). A description of Meager's consumption model
for non-zero consumption levels follows.

Let \(d_{ik} \in \{a,b\}\) denote the treatment assignment to individual
\(i\) in site \(k\), where \(k = 1,...,5.\) Let \texttt{MvN} represent
the multivariate normal distribution, and let \(I\) denotes the identity
matrix. We then have:

\begin{tcolorbox}[enhanced jigsaw, colframe=quarto-callout-note-color-frame, left=2mm, bottomtitle=1mm, leftrule=.75mm, opacitybacktitle=0.6, toptitle=1mm, arc=.35mm, rightrule=.15mm, colback=white, coltitle=black, breakable, colbacktitle=quarto-callout-note-color!10!white, toprule=.15mm, titlerule=0mm, title={A Bayesian hierarchical model}, opacityback=0, bottomrule=.15mm]

\[
y_{ik}(d_{ik}) \sim \texttt{LogNormal}(\mu_k + \zeta_k 1_b(d_{ik}),\sigma_k \lambda_k^{1_b(d_{ik})}) \text{ for } k = 1,...,5; 
\] \[
0.1\mu_k, 0.1\zeta_k, \log(\sigma_k), \log(\lambda_k) \sim \texttt{MvN}(0, 10I) \text{ for } k = 1,...,5.
\]

\end{tcolorbox}

The interpretation is that, for every site \(k\), consumption is
lognormally distributed, with log-mean \(\mu_{k}\) and log-standard
deviation \(\sigma_{k}\) for \(y_{ik}(a)\), and log-mean
\(\mu_{k} + \zeta_{k}\) and log-standard deviation
\(\sigma_{k} \lambda_{k}\) for \(y_{ik}(b)\).\footnote{This description
  is contained in the file
  \texttt{tailored-hierarchical-pdf-log-normal-1-\ tail.stan}, which can
  be found in the repository for Meager (2022), available at
  \url{https://bitbucket.org/rmeager/aggregating-distributional-treatment-effects/src/master/}.}
Let \(y_k(d)\) denote the vector of consumption in site \(k\) given
treatment \(d \in \{a,b\}\). Then, with this structure in place, I am
able to obtain a closed form solution for the egalitarian equivalent of
\(y_k(d)\):

\begin{equation}\phantomsection\label{eq-ee}{
\mathcal{EE}(y_k(d)) = e^{\mu_k + \zeta_k 1_b(d) + \frac{1}{2}(1 - \gamma) \left(\sigma_k \lambda_k^{1_b(d)}\right)^2}.
}\end{equation}

I consider here, as in Section~\ref{sec-jobcorps}, an evaluator with
inequality aversion parameter of \(\gamma = 2\). I then compute mean
treatment effects and egalitarian equivalent treatment effects using
Meager's Markov Chain Monte Carlo (MCMC) output, denoted \(\hat{\pi}\),
which contains three chains with four thousand draws per
chain,\footnote{These draws are contained in the file
  \texttt{microcredit\_consumption\_lognormal\_tailored\ \_hierarchical\_pdf\_output\_5000\_iters.RData},
  which can be found in the paper's repository reported above.} and
where I am able to use Equation~\ref{eq-ee} to calculate the egalitarian
equivalent measures in every draw.\footnote{The mean income measures can
  also be calculated in every MCMC draw using Equation~\ref{eq-ee} with
  \(\gamma = 0\).}

\begin{figure}

\centering{

\includegraphics{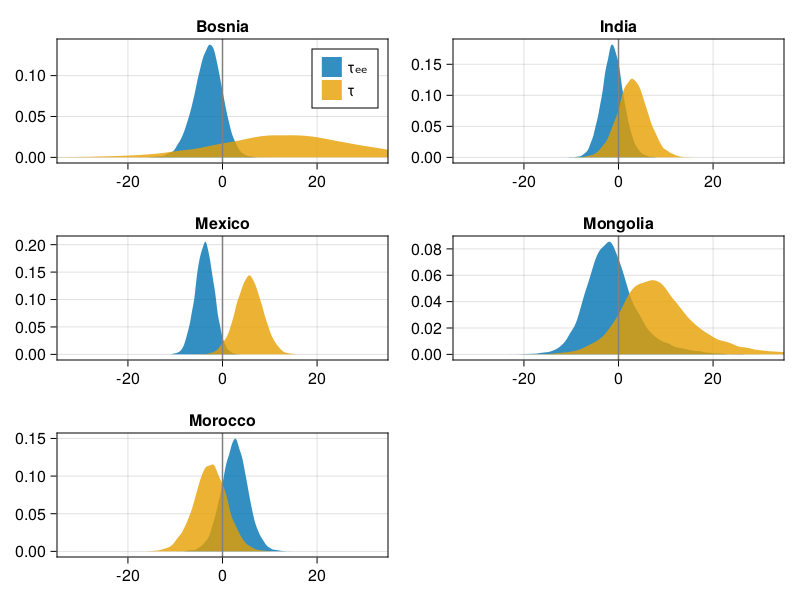}

}

\caption{\label{fig-posteriors}Posterior distributions of the mean
treatment effects (\(\tau\)) and the egalitarian equivalent treatment
effects (\(\tau_{ee}\)). All units are 2009 USD PPP per two weeks.}

\end{figure}

Figure~\ref{fig-posteriors} reports the posterior distributions of the
treatment effects \(\tau(y_k):=E[y_k(b)] - E[y_k(a)]\) and
\(\tau_{ee}(y_k):=\mathcal{EE}(y_k(b)) - \mathcal{EE}(y_k(a))\), and
Table~\ref{tbl-mean_results} reports estimates for
\(\hat{\tau}(k):=E_{\hat{\pi}}[\tau(y_k)]\) and
\(\hat{\tau}_{ee}(k) := E_{\hat{\pi}}[\tau_{ee}(y_k)]\) for the five
countries (\(k=1,...,5\)), where the expectations are computed using the
MCMC draws \(\hat{\pi}\). Table~\ref{tbl-mean_results} also reports the
posterior probabilities, \(P_{\hat{\pi}}[\tau(y_k)>0]\) and
\(P_{\hat{\pi}}[\tau_{ee}(y_k)>0]\), that the average and egalitarian
equivalent treatment effects are positive according to those draws.

\newpage

\begin{longtable}[]{@{}
  >{\centering\arraybackslash}p{(\columnwidth - 8\tabcolsep) * \real{0.2000}}
  >{\centering\arraybackslash}p{(\columnwidth - 8\tabcolsep) * \real{0.2000}}
  >{\centering\arraybackslash}p{(\columnwidth - 8\tabcolsep) * \real{0.2000}}
  >{\centering\arraybackslash}p{(\columnwidth - 8\tabcolsep) * \real{0.2000}}
  >{\centering\arraybackslash}p{(\columnwidth - 8\tabcolsep) * \real{0.2000}}@{}}
\caption{Microcredit treatment effect Bayesian
estimates}\label{tbl-mean_results}\tabularnewline
\toprule\noalign{}
\begin{minipage}[b]{\linewidth}\centering
\end{minipage} & \begin{minipage}[b]{\linewidth}\centering
\(E_{\hat{\pi}}[\tau]\)
\end{minipage} & \begin{minipage}[b]{\linewidth}\centering
\(P_{\hat{\pi}}[\tau>0]\)
\end{minipage} & \begin{minipage}[b]{\linewidth}\centering
\(E_{\hat{\pi}}[\tau_{ee}]\)
\end{minipage} & \begin{minipage}[b]{\linewidth}\centering
\(P_{\hat{\pi}}[\tau_{ee}>0]\)
\end{minipage} \\
\midrule\noalign{}
\endfirsthead
\toprule\noalign{}
\begin{minipage}[b]{\linewidth}\centering
\end{minipage} & \begin{minipage}[b]{\linewidth}\centering
\(E_{\hat{\pi}}[\tau]\)
\end{minipage} & \begin{minipage}[b]{\linewidth}\centering
\(P_{\hat{\pi}}[\tau>0]\)
\end{minipage} & \begin{minipage}[b]{\linewidth}\centering
\(E_{\hat{\pi}}[\tau_{ee}]\)
\end{minipage} & \begin{minipage}[b]{\linewidth}\centering
\(P_{\hat{\pi}}[\tau_{ee}>0]\)
\end{minipage} \\
\midrule\noalign{}
\endhead
\bottomrule\noalign{}
\endlastfoot
Bosnia & 13.82 & 82.5\% & -3.11 & 14.3\% \\
India & 3.00 & 83.1\% & -1.20 & 29.1\% \\
Mexico & 5.65 & 97.5\% & -3.76 & 2.7\% \\
Mongolia & 8.58 & 88.1\% & -1.35 & 35.0\% \\
Morocco & -2.57 & 22.7\% & 2.55 & 82.4\% \\
\end{longtable}

Using the results from Table~\ref{tbl-mean_results} one reaches the
conclusion that incorporating inequality considerations into the
analysis can plausibly reverse the policy recommendation one would make
if one focuses solely on what happens on average across the distribution
of outcomes, and sharpen the recommendations one would make if one
focuses on the quantile treatment effects without further aggregation.
To see this, Table 5 presents Meager (2022)'s Bayesian quantile
treatment effect estimates on consumption.

We see that the treatment effects on mean consumption reported in
Table~\ref{tbl-mean_results} are driven by large changes that take place
at the top quantiles of the distribution of consumption in Table 5,
whereas the egalitarian equivalent treatment effect estimates reported
in Table~\ref{tbl-mean_results} are driven by small changes that take
place at the bottom quantiles of the distribution of consumption in
Table 5. The egalitarian equivalent treatment effect approach executes a
principled aggregation of the treatment effect heterogeneity across
quantiles, according to the attitudes towards inequality of the
evaluator, and the aggregation tools developed through Theorem 3.1 in
Zambrano (2024) and Theorem~\ref{thm-AB} in the present paper.

\begin{remark}
For comparison, consider how one would analyze the data generated above
using conventional tools from the randomized evaluation literature. The
most straightforward analysis would be a differences in means
comparison. In this case, the point estimates (with standard errors in
parentheses) of the differences in means \(E[y(b)] - E[y(a)]\) for the
five sites are (in 2009 USD PPP per two weeks): Bosnia -1.59 (14.14),
India 4.55 (3.85) (India), Mexico 5.51 (2.90), Mongolia 50.45 (15.67)
and Morocco -2.93 (4.26). One would reject the hypothesis that the
treatments have the same effect on average income at the 5\% level in
the case of Mongolia, and would not reject the hypothesis in the other
four countries.\footnote{This set of results is illustrated in Figure 3,
  Panel D, in Meager (2019).}
\end{remark}

\begin{table}

\caption{\label{tbl-meager_table}Bayesian quantile treatment effects on
consumption (from Table 1 in Meager (2022))}

\centering{

\includegraphics{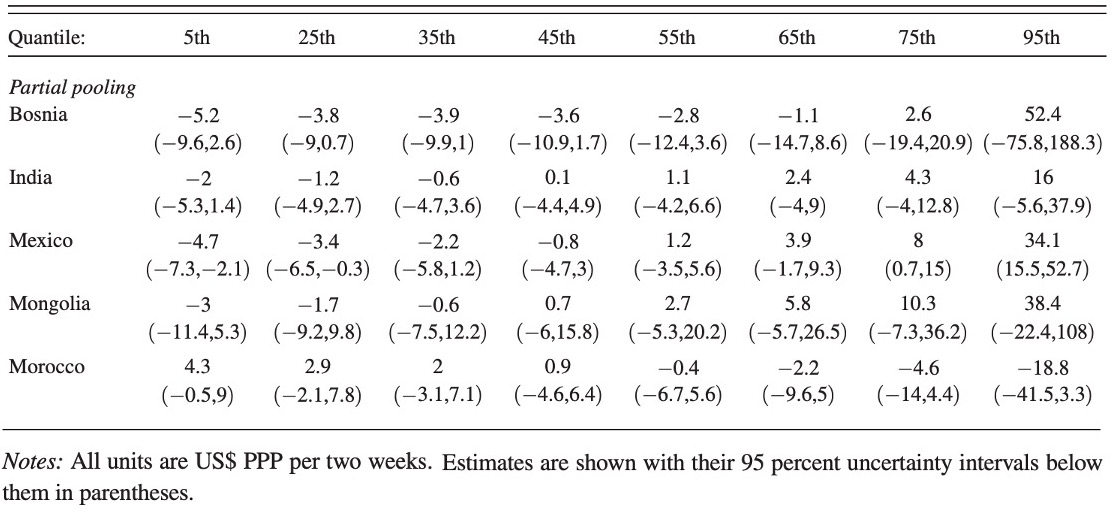}

}

\end{table}%

\section{Large Sample Analysis in the Limit of Experiments
Framework}\label{large-sample-analysis-in-the-limit-of-experiments-framework}

Given a statistical treatment rule, and before the realization of the
sample data, the profile
\((ee^{1}(\delta(Z^t)),...,ee^{m}(\delta(Z^t)))\) is a random vector. In
order to study this profile, \emph{ex ante}, we therefore need to extend
the preferences of the evaluator over \(\Upsilon_{1}\) to account for
the additional sources of uncertanity (sampling uncertanity) that the
evaluator faces. As in Manski (2004) (and following Wald (1971)), one
can measure the performance of \(\delta(\cdot)\) in state \(s\) by its
expected welfare across samples, which in this case means
\(E_{s} \left[ee^{s}(\delta(Z^t))\right]\)
\(:=\int{ee^{s}(\delta(Z^{t}))dP_{s}^{t}(Z^{t})}\). Not knowing the true
state, the evaluator then assesses \(\delta(\cdot)\) by the
state-dependent expected egalitarian equivalent vector
\(\left(E_{s}\left[ee^{s}(\delta(Z^{t}))\right]\right)_{s \in S}\). Let
\(\Gamma\) denote the set of statistical treatment rules and and assume
that \(S\) is an open subset of \(\mathbb{R}^{J}\), \(J>0\). Let \(\pi\)
here denote a prior measure on \(S\).

The \emph{ex-ante} versions of Equation~\ref{eq-bayesian},
Equation~\ref{eq-maximin} and Equation~\ref{eq-mmr} in this setting are
\begin{equation}\phantomsection\label{eq-bayesianante}{
\sup_{\delta(\cdot) \in \Gamma} E_{\pi}\left[E_{s}\left[ee^{s}(\delta(Z^{t}))\right] \right] 
}\end{equation} \begin{equation}\phantomsection\label{eq-maximinante}{
\sup_{\delta(\cdot) \in \Gamma} \inf_{s \in S} E_{s}\left[ee^{s}(\delta(Z^{t}))\right]  
}\end{equation} \begin{equation}\phantomsection\label{eq-mmrante}{
\inf_{\delta(\cdot) \in \Gamma} \sup_{s \in S} \left[ \max\{ee^{s}(a), ee^{s}(b)\} - E_{s}\left[ee^{s}(\delta(Z^{t}))\right]\right] 
}\end{equation}

I focus below on obtaining asymptotically optimal solutions to
Equation~\ref{eq-bayesianante} and Equation~\ref{eq-mmrante}. To do so,
I consider a sequence of experiments \(\{P^{t}_{s}, s \in S\}\) as the
sample size \(t\) grows.

Let \(s_0\) be such that \(\tau_{ee}(s_0) = 0\), noting that the
presence of sampling uncertainty may make it difficult to distinguish
between treatments in terms of their egalitarian equivalents if the true
state is close to \(s_0\). Following Hirano and Porter (2009), I
reparametrize the state space and consider parametric sequences of the
form \(s_0 + \frac{h}{\sqrt{t}}\) for \(h \in \mathbb{R}^{J}\). The
intuition behind the choice is the following: for \(\ddot{s}\) such that
\(ee^{\ddot{s}}(a) \ne ee^{\ddot{s}}(b)\), the treatment that is better
at \(\ddot{s}\) will be better for all local alternatives
\(\ddot{s} + \frac{h}{\sqrt{t}}\) asymptotically. Therefore, a local
reparametrization centered around \(s_0\) ensure that we are looking a
the cases where it is hardest to determine which is the better treatment
even as the sample size grows large. In what follows, I will employ the
following assumptions, which are standard in the limit of experiments
framework (Le Cam (2012), Van der Vaart (1998), Hirano and Porter
(2020)):

\begin{itemize}
\item
  \textbf{DQM\('\)}. A sequence of experiments
  \(\{P^{t}_{s}, s \in S\}\) satisfies \textbf{DQM\('\)} at \(s_0\) if
  there exists a score function
  \(\mathscr{s}:\mathcal{Z}\rightarrow \mathbb{R}^{J}\), with
  nonsingular Fisher information matrix
  \(I_{0} = E_{s_0}[\mathscr{s}\mathscr{s}']\), such that
  \(\int [dP_{s_0+h}^{\frac{1}{2}}(Z)-\)
  \(dP_{s_0}^{\frac{1}{2}}(Z) - \frac{1}{2} h'\mathscr{s}(Z)\)
  \(dP_{s_0}^{\frac{1}{2}}(Z)]^2 = \mathscr{o}(||h||^{2})\) as
  \(h \rightarrow 0.\)
\item
  \textbf{C}. A sequence of statistical treatment rules \(\delta_{t}\)
  in the experiments \(\{P^{t}_{s}, s \in S\}\) satisfies \textbf{C} if
  \(E_{s_0 + \frac{h}{\sqrt{t}}}\left[\delta_{t}(Z^{t})\right]\) has a
  well defined limit for all \(h\in \mathbb{R}^J\).
\item
  \textbf{h-BRE}. An estimator \(\hat{s_{t}}\) of \(s\) satisfies
  \textbf{h-BRE} if, for all \(h \in \mathbb{R}^J\),
  \(\sqrt{t}\left(\hat{s}_{t} - s_0 - \frac{h}{\sqrt{t}} \right) \leadsto  N(0,I_{0}^{-1})\)
  under the sequence of probability measures
  \(P^{t}_{s_0 + \frac{h}{\sqrt{t}}}\).
\item
  \textbf{L}. A prior measure \(\pi\) on \(S\) satisfies \textbf{L} if
  it admits a density with respect to Lebesgue measure that is
  continuous and positive at \(s_0\).
\end{itemize}

The result below is a consequence of Proposition 3.1 and Theorems 3.2,
3.4 and 3.5 in Hirano and Porter (2009):

\begin{theorem}[]\protect\hypertarget{thm-HP}{}\label{thm-HP}

Assume that \(s_0 \in S\), the sequence of experiments
\(\{P^{t}_{s}, s \in S\}\) satisfies \textbf{DQM\('\)}, the sequence of
statistical treatment rules \(\delta_{t}\) in the experiments
\(\{P^{t}_{s}, s \in S\}\) satisfies \textbf{C}, the prior measure
\(\Pi\) on \(S\) satisfies \textbf{L} and the estimator \(\hat{s}_t\) of
\(s\) satisfies \textbf{h-BRE}. Then the feasible statistical treatment
rule \(\delta^*_t(Z^{t}) = 1 \left(\tau_{ee}(\hat{s}_{t}) > 0 \right)\)
is asypmtotically Bayes and minimax regret optimal.

\end{theorem}

The interpretation is that, under the structure provided by the
assumptions in Theorem~\ref{thm-HP}, one can act as if all relevant
uncertanity has been resolved in large samples, and make a decision that
would be optimal if the point estimate \(\hat{s}_t\) of \(s\) were
accurate.

\subsection{Application: Microcredits Reexamined}\label{sec-HP_meager}

Below I revisit the application in Section~\ref{sec-meager} under the
assumptions behind Theorem~\ref{thm-HP}, and taking the sample sizes of
the five RCTs to be large. From Meager's MCMC output we can obtain
Bayesian point estimates of the profile
\(\left(\mu_k, \zeta_k, \sigma_k, \lambda_k\right)\) from the Bayesian
hierarchical model from Section~\ref{sec-meager}, and for the five
countries \((k = 1, ..., 5)\). Plugging these estimates into
Equation~\ref{eq-ee}, we can then obtain estimates of
\(\mathcal{EE}(y_k(d))\), \(d \in \{a,b\}\), and then of
\(\tau_{ee}(s)\). Table~\ref{tbl-HP_meager} below reports the results
from this analysis.

\begin{longtable}[]{@{}
  >{\centering\arraybackslash}p{(\columnwidth - 8\tabcolsep) * \real{0.2000}}
  >{\centering\arraybackslash}p{(\columnwidth - 8\tabcolsep) * \real{0.2000}}
  >{\centering\arraybackslash}p{(\columnwidth - 8\tabcolsep) * \real{0.2000}}
  >{\centering\arraybackslash}p{(\columnwidth - 8\tabcolsep) * \real{0.2000}}
  >{\centering\arraybackslash}p{(\columnwidth - 8\tabcolsep) * \real{0.2000}}@{}}
\caption{Microcredit treatment effect large sample Bayesian
estimates}\label{tbl-HP_meager}\tabularnewline
\toprule\noalign{}
\begin{minipage}[b]{\linewidth}\centering
\end{minipage} & \begin{minipage}[b]{\linewidth}\centering
\(N_a\)
\end{minipage} & \begin{minipage}[b]{\linewidth}\centering
\(N_b\)
\end{minipage} & \begin{minipage}[b]{\linewidth}\centering
\(\tau\left(\hat{\mu}, \hat{\zeta}, \hat{\sigma}, \hat{\lambda}\right)\)
\end{minipage} & \begin{minipage}[b]{\linewidth}\centering
\(\tau_{ee}\left(\hat{\mu}, \hat{\zeta}, \hat{\sigma}, \hat{\lambda}\right)\)
\end{minipage} \\
\midrule\noalign{}
\endfirsthead
\toprule\noalign{}
\begin{minipage}[b]{\linewidth}\centering
\end{minipage} & \begin{minipage}[b]{\linewidth}\centering
\(N_a\)
\end{minipage} & \begin{minipage}[b]{\linewidth}\centering
\(N_b\)
\end{minipage} & \begin{minipage}[b]{\linewidth}\centering
\(\tau\left(\hat{\mu}, \hat{\zeta}, \hat{\sigma}, \hat{\lambda}\right)\)
\end{minipage} & \begin{minipage}[b]{\linewidth}\centering
\(\tau_{ee}\left(\hat{\mu}, \hat{\zeta}, \hat{\sigma}, \hat{\lambda}\right)\)
\end{minipage} \\
\midrule\noalign{}
\endhead
\bottomrule\noalign{}
\endlastfoot
Bosnia & 427 & 520 & 13.76 & -3.11 \\
India & 3247 & 3579 & 3.00 & -1.20 \\
Mexico & 8296 & 8260 & 5.65 & -3.76 \\
Mongolia & 260 & 701 & 8.72 & -1.35 \\
Morocco & 2771 & 2716 & -2.57 & 2.55 \\
\end{longtable}

These results are essentially identical to those reported in
Table~\ref{tbl-mean_results} and the conclusions one reaches about the
optimal treatment assignment for each of these countries are therefore
the same.

\section{Summary}\label{summary}

My aim with this paper is to contribute towards the development of an
integrated theory of how account for inequality in the distribution of
treatment effects in experimental and observational settings. To adopt
the tools developed above in an applied setting, the recommended
workflow would be as follows: First, use the results in Fleurbaey and
Zambrano (2024) to help determine which SP under certainty to bring into
the analysis. Second, use Theorem 3.1 in Zambrano (2024) to determine
how to extend that SP to a world where risk, uncertainty or ambiguity
play a prominent role. Third, use the tools in Flores, Kairy, and
Zambrano (2024) to properly estimate egalitarian equivalent treatment
effects, and bounds, together with corresponding uncertainty estimates
for these. Fourth, and last, use the results in Section~\ref{sec-point},
Theorem~\ref{thm-AB} and Theorem~\ref{thm-HP}, to identify the optimal
treatment assignment rule needed for the statistical decision problem
that an inequality sensitive evaluator may want to solve.

An important missing ingredient in the present analysis is that I make
no use of covariate information. Introducing covariates is known to help
the evaluator obtain better bounds on treatment effects (as in Lee
(2009) and Semenova (2023)), but it also opens up the question as to
whether the evaluator wishes to account solely for the inequality
generated by a treatment conditional on covariates, or also incorporate
the inequality generated by a treatment across the multiple values that
the covariates may take. Taking a look at both of these important issues
is left for future work.

\section{Appendix}\label{appendix}

\subsection{\texorpdfstring{Proof of
Theorem~\ref{thm-AB}}{Proof of Theorem~}}\label{proof-of-thm-ab}

First, consider the solution to the Bayesian decision problem.

Since \(f^{-1}\) is strictly convex, for each \(s \in S\) and
\(\delta(P_s) \in (0,1)\):

\[f^{-1}\left[\delta(P_s)f(ee^{s}(b))+(1-\delta(P_s))f(ee^{s}(a))\right] < \delta(P_s) \cdot ee^{s}(b) + (1-\delta(P_s)) \cdot ee^{s}(a).
\]

Take the expectation of both sides:

\[ 
\begin{aligned}
& E_\pi\left[f^{-1}\left[\delta(P_s)f(ee^{s}(b))+(1-\delta(P_s))f(ee^{s}(a))\right]|S(P_s)\right] < 
\\
& \delta(P_s) \cdot E_\pi\left[ee^{s}(b)|S(P_s)\right] + (1-\delta(P_s)) \cdot E_\pi\left[ee^{s}(a)|S(P_s)\right].
\end{aligned}
\]

It follows that

\begin{equation}\phantomsection\label{eq-propAB}{
\begin{aligned}
&E_\pi\left[f^{-1}\left[\delta(P_s)f(ee^{s}(b))+(1-\delta(P_s))f(ee^{s}(a))\right]|S(P_s)\right] <
\\
& \max\{E_\pi\left[ee^{s}(b)|S(P_s)\right], E_\pi\left[ee^{s}(a)|S(P_s)\right]\}.
\end{aligned}
}\end{equation}

Therefore, if \(E_{\pi}\left[\tau_{ee}(s)|S(P_s)\right]>0\), then
\(E_\pi\left[ee^{s}(b)|S(P_s)\right]> E_\pi\left[ee^{s}(a)|S(P_s)\right]\)
and Equation~\ref{eq-propAB} implies that \(\delta^{B}(P_s) = 1\). On
the other hand, if \(E_{\pi}\left[\tau_{ee}(s)|S(P_s)\right]\le 0\),
then
\(E_\pi\left[ee^{s}(b)|S(P_s)\right]\le E_\pi\left[ee^{s}(a)|S(P_s)\right]\)
and Equation~\ref{eq-propAB}, together with our tie-breaking rule,
implies that \(\delta^{B}(P_s) = 0\).

Now, consider the solution to the maximin decision problem.

Since \(s_w\in S(P_s)\), and \(s_w\) is a worst state for both
treatments, the problem amounts to maximizing

\[ee^{s_w}(\delta(P_s)) = f^{-1}\left[\delta(P_s)f(ee^{s_w}(b))+(1-\delta(P_s))f(ee^{s_w}(a))\right],\]

which is equivalent to maximizing
\(f(ee^{s_w}(a)) + \delta(P_s) \left(f(ee^{s_w}(b))-f(ee^{s_w}(a))\right)\).

Therefore, if \(ee^{s_w}(b)> ee^{s_w}(a)\) then
\(f(ee^{s_w}(b))-f(ee^{s_w}(a))>0\) and \(\delta^{M}(P_s) = 1\). On the
other hand, if \(ee^{s_w}(b)\le ee^{s_w}(a)\) then
\(f(ee^{s_w}(b))-f(ee^{s_w}(a)) \le 0\) and this fact, together with our
tie-breaking rule, implies that \(\delta^{M}(P_s) = 0\).

Last, consider the solution to the minimax regret decision problem.

Consider states \(s'\in S(P_s)\) such that \(ee^{s'}(a)> ee^{s'}(b)\).
Let \(R_a(\delta, s'):=ee^{s'}(a) - ee^{s'}(\delta)\) Notice that
\(R_a(\delta, s')\) is strictly increasing in \(\delta\),
\(R_a(0, s')=0\) and \(R_a(1, s') = ee^{s'}(a) - ee^{s'}(b)>0\). Notice
also that, for fixed \(\delta\), \(R_a(\delta, s')\) is greatest when
\(s'=s_a\).

Now consider states \(s''\in S(P_s)\) such that
\(ee^{s''}(b)> ee^{s''}(a)\). Let
\(R_b(\delta, s''):=ee^{s''}(b) - ee^{s''}(\delta)\). Notice that
\(R_b(\delta, s'')\) is strictly decreasing in \(\delta\),
\(R_b(0, s'') = ee^{s''}(b) - ee^{s''}(a)>0\) and \(R_b(1, s'')=0\).
Notice also that, for fixed \(\delta\), \(R_b(\delta, s'')\) is greatest
when \(s''=s_b\).

The objective function to be minimized can therefore be written as

\[
\max_{s \in \{s_a,s_b\}} \left[ \max \{ee^{s}(a), ee^{s}(b)\} - ee^{s}(\delta(P_s))\right] 
\]

which is continuous in \(\delta\) and therefore attains a minimum in the
interval {[}0,1{]}.

Now consider the function
\(H(\delta) = R_b(\delta, s_b)-R_a(\delta, s_a)\). It readily follows
that that \(H(0)>0\) and \(H(1)<0\). Since \(H\) is continuous, there is
\(\delta^{R}(P_s) \in (0,1)\) such that \(H(\delta^{R}(P_s))=0\).

Given all this, notice that \(\delta <\delta^{R}(P_s)\) cannot be
minimax regret optimal, since in this case worst regret is given by
\(R_b(\delta, s_b)>0\), which can be lowered by slightly increasing
\(\delta\), given that \(R_b(\delta, s_b)\) is strictly decreasing.
Similarly, notice that \(\delta > \delta^{R}(P_s)\) cannot be minimax
regret optimal, since in this case worst regret is given by
\(R_a(\delta, s_a)>0\), which can be lowered by slightly decreasing
\(\delta\), given that \(R_a(\delta, s_a)\) is strictly increasing.
Therefore, by the definition of \(H(\delta)\), \(\delta^{R}(P_s)\) such
that

\[ee^{s_{a}}(a) - ee^{s_{a}}(\delta^{R}(P_s)) = ee^{s_{b}}(b) - ee^{s_{b}}(\delta^{R}(P_s)).\]

is minimax regret optimal, which is what we wanted to show.

\subsection{\texorpdfstring{Proof of
Proposition~\ref{prp-IN}}{Proof of Proposition~}}\label{proof-of-prp-in}

Assume the \(IN\) evaluator accepts the innovation. Then, from Manski
(2004), p.~1228, we know that \(E_{\pi}[p(b)] > p(a)\) and
\(\delta_{in} = 1\). We want to show that
\(E_\pi[f^{-1}(p^s(b))] > E_\pi[f^{-1}(p(a) + \delta \cdot (p^s(b) - p(a)))]\)
for \(\delta \in [0,1)\). To see this, notice first that, for
\(\delta = 0\), we have \(E_\pi[f^{-1}(p(a))] < E_\pi[f^{-1}(p^s(b))]\),
which is true because \(E_\pi[p^s(b)] > p(a)\) and \(f^{-1}\) is
strictly increasing. Now, since \(f^{-1}\) is strictly convex, for each
\(s \in S\) and \(\delta \in (0,1)\):

\[f^{-1}(\delta \cdot p^s(b) + (1-\delta) \cdot p(a)) < \delta \cdot f^{-1}(p^s(b)) + (1-\delta) \cdot f^{-1}(p(a)).\]

Take the expectation of both sides:
\[E_\pi[f^{-1}(p(a) + \delta \cdot (p^s(b) - p(a)))] < E_\pi[\delta \cdot f^{-1}(p^s(b)) + (1-\delta) \cdot f^{-1}(p(a))],\]

and notice that
\[E_\pi[\delta \cdot f^{-1}(p^s(b)) + (1-\delta) \cdot f^{-1}(p(a))] = \delta \cdot E_\pi[f^{-1}(p^s(b))] + (1-\delta) \cdot f^{-1}(p(a)),\]

which means that

\begin{equation}\phantomsection\label{eq-prop310}{E_\pi[f^{-1}(p(a) + \delta \cdot (p^s(b) - p(a)))] < \delta \cdot E_\pi[f^{-1}(p^s(b))] + (1-\delta) \cdot f^{-1}(p(a)).}\end{equation}

Since \(E_\pi[p^s(b)] > p(a)\), we obtain that
\(f^{-1}(E_\pi[p^s(b)]) > f^{-1}(p(a))\). Then, by Jensen's inequality,

\begin{equation}\phantomsection\label{eq-prop311}{E_\pi[f^{-1}(p^s(b))] > f^{-1}(E_\pi[p^s(b)]) > f^{-1}(p(a))}\end{equation}

Combining Equation~\ref{eq-prop310} and Equation~\ref{eq-prop311}, we
obtain

\[
\begin{aligned}
& E_\pi[f^{-1}(p(a) + \delta \cdot (p^s(b) - p(a)))] <
\\
 & \delta \cdot E_\pi[f^{-1}(p^s(b))] + (1-\delta) \cdot E_\pi[f^{-1}(p^s(b))] = E_\pi[f^{-1}(p^s(b))],
\end{aligned}
 \]

Therefore, we have shown that
\(E_\pi[f^{-1}(p^s(b))] > E_\pi[f^{-1}(p(a) + \delta \cdot (p^s(b) - p(a)))]\)
for all \(\delta \in [0,1)\), which completes the proof.

\subsection{\texorpdfstring{Proof of
Proposition~\ref{prp-IA}}{Proof of Proposition~}}\label{proof-of-prp-ia}

The solution to the \(IN\) minimax regret problem is:
\(\delta^{*}_{in} = 1 - p(a)\). (Stoye (2009), p.~73). The solution to
the \(IA\) minimax regret problem is \(\delta^*_{ia}\) such that

\begin{equation}\phantomsection\label{eq-IA_proof}{ \bar{p} - f^{-1}(p(a)+(1-p(a))\delta^*_{ia}) =f^{-1}(p(a)) - f^{-1}(p(a)(1-\delta^*_{ia}))}\end{equation}

for all \(p(a) \in (0,1).\)

Notice that \(\delta = 0\) cannot be the solution, since

\[ \bar{p} - f^{-1}(p(a)) > f^{-1}(p(a)) - f^{-1}(p(a)).\]

Similarly, \(\delta = 1\) cannot be the solution since

\[  0 = \bar{p} - f^{-1}(1) <f^{-1}(p(a)) - f^{-1}(0) = f^{-1}(p(a)) - \underline{p}  \]

Therefore \(\delta^*_{ia} \in (0,1).\)

Consider the following function of two variables:

\[
g(y, b) = f^{-1}(y + b) - f^{-1}(y), \quad b > 0, \quad y \in (0,1).
\]

Because \(f\) is strictly increasing and strictly concave, the
derivatives of \(g\) with respect to both \(y\) and \(b\) are positive
for \(b > 0\) and \(y \in (0,1)\):

\[
g'_b = f^{-1'}(y + b) > 0,
\]

\[
g'_y = f^{-1'}(y + b) - f^{-1'}(y) > 0.
\]

Now consider a level curve in the \((y, b)\) plane on which the function
\(g\) is constant:

\[
g(y, b) = \text{const}.
\]

Differentiating, we get:

\[
f'_y \, dy + f'_b \, db = 0.
\]

From the above equations, on the level curve:

\[
\frac{dy}{db} < 0.
\]

This implies:

\[
g(y_1, b_1) = g(y_2, b_2) \text{ and } y_1 < y_2 \Rightarrow b_1 > b_2.
\]

Now, let:

\[
y_1 = a > y_2 = a(1-\delta^*_{ia}), \quad b_1 = 1 - a, \quad b_2 = \delta^*_{ia}.
\]

From these equations:

\[
1 - a < \delta^*_{ia}.
\]

\subsection{\texorpdfstring{Proof of
Theorem~\ref{thm-HP}}{Proof of Theorem~}}\label{proof-of-thm-hp}

Below I follow the argument and presentation in Hirano and Porter
(2020), pp.~333-334, adapted to the present setting. Assume that the
sequence of experiments \(\{P^{t}_{s}, s \in S\}\) satisfies
\textbf{DQM\('\)} at \(s_0\) and that the sequence of statistical
treatment rules \(\delta_{t}\) in the experiments
\(\{P^{t}_{s}, s \in S\}\) satisfies \textbf{C}. Under these
assumptions, we know from Proposition 3.1 in Hirano and Porter (2009)
that there is a statistical treatment rule
\(\delta:\mathbb{R}^J \rightarrow [0,1]\) such that

\[
\lim_{t \rightarrow \infty} E_{s_0 + \frac{h}{\sqrt{t}}}\left[\delta_{t}(Z^{t})\right]=\int \delta(\xi)dN(\xi|h,I^{-1}_0)
\] for all \(h\in \mathbb{R}^J\), where \(N(\xi|h,I^{-1}_0)\), a
Gaussian distribution with mean \(h\) and variance \(I^{-1}_0\), is the
limit experiment for the original problem. In this limit experiment, one
observes a single draw from \(Z \sim N(h,I^{-1}_0)\) and makes decisions
based on this draw. Since \(\tau_{ee}(s_0) = 0\), it follows that
\(\sqrt{t} \; \tau_{ee}\left(s_0 + \frac{h}{\sqrt{t}}\right) \rightarrow \nabla \tau_{ee}(s_0)' \cdot h\)
as \(t \rightarrow \infty\). Hirano and Porter (2009) (pp.~1691 and
1693) show that the solutions to the Bayesian and the minimax
statistical decision problems in the limit experiment are the same and
are based on the known linear function \(\nabla \tau_{ee}(s_0)'\cdot Z\)
of the estimator \(Z\) of \(h\). In particular, the solution,
\(\delta^*\), to both problems in the limit experiment is given by:

\[\delta^*(Z) = 1\left(\nabla \tau_{ee}(s_0)'\cdot Z > 0 \right).\]

If one also assumes that the estimator \(\hat{s_{t}}\) of \(s\)
satisfies \textbf{h-BRE}, the feasible statistical treatment rule

\[\delta^*_t(Z^{t}) = 1 \left(\tau_{ee}(\hat{s}_{t}) > 0 \right)\]

will have limiting distributions that match
\(1\left(\nabla \tau_{ee}(s_0)'\cdot Z > 0 \right)\) (Hirano and Porter
(2009), p.~1692). Theorem 3.5 in Hirano and Porter (2009) then shows
that \(1 \left(\tau_{ee}(\hat{s}_{t}) > 0 \right)\) is asypmtotically
minimax regret optimal. If, in addition, one assumes that the prior
measure \(\pi\) on \(S\) satisfies \textbf{L}, Theorem 3.2 in Hirano and
Porter (2009) shows that \(1 \left(\tau_{ee}(\hat{s}_{t}) > 0 \right)\),
is asymptotically Bayes optimal as well.

\section*{References}\label{references}
\addcontentsline{toc}{section}{References}

\phantomsection\label{refs}


\begin{CSLReferences}{1}{0}
\bibitem[\citeproctext]{ref-adler_theory_2022}
Adler, Matthew D. 2022. {``Theory of {Prioritarianism}.''} In
\emph{Prioritarianism in {Practice}}, edited by Matthew D. Adler and Ole
F. Norheim. Cambridge, United Kingdom ; New York, NY: Cambridge
University Press.

\bibitem[\citeproctext]{ref-athey_policy_2021}
Athey, Susan, and Stefan Wager. 2021. {``Policy {Learning} {With}
{Observational} {Data}.''} \emph{Econometrica} 89 (1): 133--61.
\url{https://doi.org/10.3982/ECTA15732}.

\bibitem[\citeproctext]{ref-atkinson_measurement_1970}
Atkinson, Anthony B. 1970. {``On the Measurement of Inequality.''}
\emph{Journal of Economic Theory} 2 (3): 244--63.
\url{https://doi.org/10.1016/0022-0531(70)90039-6}.

\bibitem[\citeproctext]{ref-beygelzimer_offset_2009}
Beygelzimer, Alina, and John Langford. 2009. {``The Offset Tree for
Learning with Partial Labels.''} In \emph{Proceedings of the 15th {ACM}
{SIGKDD} International Conference on {Knowledge} Discovery and Data
Mining}, 129--38. Paris France: ACM.
\url{https://doi.org/10.1145/1557019.1557040}.

\bibitem[\citeproctext]{ref-blackorby_measures_1978}
Blackorby, Charles, and David Donaldson. 1978. {``Measures of Relative
Equality and Their Meaning in Terms of Social Welfare.''} \emph{Journal
of Economic Theory} 18 (1): 59--80.
\url{https://doi.org/10.1016/0022-0531(78)90042-X}.

\bibitem[\citeproctext]{ref-chen_bounds_2015}
Chen, Xuan, and Carlos A. Flores. 2015. {``Bounds on {Treatment}
{Effects} in the {Presence} of {Sample} {Selection} and {Noncompliance}:
{The} {Wage} {Effects} of {Job} {Corps}.''} \emph{Journal of Business \&
Economic Statistics} 33 (4): 523--40.
\url{https://www.jstor.org/stable/43701561}.

\bibitem[\citeproctext]{ref-dehejia_program_2005}
Dehejia, Rajeev H. 2005. {``Program Evaluation as a Decision Problem.''}
\emph{Journal of Econometrics} 125 (1-2): 141--73.
\url{https://doi.org/10.1016/j.jeconom.2004.04.006}.

\bibitem[\citeproctext]{ref-dudik_doubly_2011}
Dudík, Miroslav, John Langford, and Lihong Li. 2011. {``Doubly Robust
Policy Evaluation and Learning.''} In \emph{Proceedings of the 28th
{International} {Conference} on {International} {Conference} on
{Machine} {Learning}}, 1097--1104. {ICML}'11. Madison, WI, USA:
Omnipress.

\bibitem[\citeproctext]{ref-fleurbaey_assessing_2010}
Fleurbaey, Marc. 2010. {``Assessing {Risky} {Social} {Situations}.''}
\emph{Journal of Political Economy} 118 (4): 649--80.
\url{https://doi.org/10.1086/656513}.

\bibitem[\citeproctext]{ref-fleurbaey_protected_2024}
Fleurbaey, Marc, and Eduardo Zambrano. 2024. {``Protected {Income} and
{Inequality} {Aversion}.''} arXiv.
\url{https://doi.org/10.48550/arXiv.2408.04814}.

\bibitem[\citeproctext]{ref-flores_egalitarian_2024}
Flores, Carlos, Michelle Kairy, and Eduardo Zambrano. 2024.
{``Egalitarian {Equivalent} {Treatment} {Effect} {Estimation}.''}

\bibitem[\citeproctext]{ref-hirano_asymptotics_2009}
Hirano, Keisuke, and Jack Porter. 2009. {``Asymptotics for {Statistical}
{Treatment} {Rules}.''} \emph{Econometrica} 77 (5): 1683--1701.
\url{https://doi.org/10.3982/ECTA6630}.

\bibitem[\citeproctext]{ref-hirano_asymptotic_2020}
Hirano, Keisuke, and Jack R. Porter. 2020. {``Asymptotic Analysis of
Statistical Decision Rules in Econometrics*.''} In \emph{Handbook of
{Econometrics}}, edited by Steven N. Durlauf, Lars Peter Hansen, James
J. Heckman, and Rosa L. Matzkin, 7:283--354. Handbook of {Econometrics},
{Volume} {7A}. Elsevier.
\url{https://doi.org/10.1016/bs.hoe.2020.09.001}.

\bibitem[\citeproctext]{ref-horowitz_nonparametric_2000}
Horowitz, Joel L., and Charles F. Manski. 2000. {``Nonparametric
{Analysis} of {Randomized} {Experiments} with {Missing} {Covariate} and
{Outcome} {Data}.''} \emph{Journal of the American Statistical
Association} 95 (449): 77--84. \url{https://doi.org/10.2307/2669526}.

\bibitem[\citeproctext]{ref-jiang_doubly_2016}
Jiang, Nan, and Lihong Li. 2016. {``Doubly {Robust} {Off}-Policy {Value}
{Evaluation} for {Reinforcement} {Learning}.''} In \emph{Proceedings of
{The} 33rd {International} {Conference} on {Machine} {Learning}},
652--61. PMLR. \url{https://proceedings.mlr.press/v48/jiang16.html}.

\bibitem[\citeproctext]{ref-kallus_policy_2018}
Kallus, Nathan, and Angela Zhou. 2018. {``Policy {Evaluation} and
{Optimization} with {Continuous} {Treatments}.''} arXiv.
\url{https://doi.org/10.48550/arXiv.1802.06037}.

\bibitem[\citeproctext]{ref-kitagawa_treatment_2024}
Kitagawa, Toru, Sokbae Lee, and Chen Qiu. 2024. {``Treatment {Choice}
with {Nonlinear} {Regret}.''} arXiv.
\url{https://doi.org/10.48550/arXiv.2205.08586}.

\bibitem[\citeproctext]{ref-kitagawa_who_2018}
Kitagawa, Toru, and Aleksey Tetenov. 2018. {``Who {Should} {Be}
{Treated}? {Empirical} {Welfare} {Maximization} {Methods} for
{Treatment} {Choice}.''} \emph{Econometrica} 86 (2): 591--616.
\url{https://doi.org/10.3982/ECTA13288}.

\bibitem[\citeproctext]{ref-kitagawa_equality-minded_2021}
---------. 2021. {``Equality-{Minded} {Treatment} {Choice}.''}
\emph{Journal of Business \& Economic Statistics} 39 (2): 561--74.
\url{https://doi.org/10.1080/07350015.2019.1688664}.

\bibitem[\citeproctext]{ref-lecam_asymptotic_2012}
Le Cam, Lucien. 2012. \emph{Asymptotic {Methods} in {Statistical}
{Decision} {Theory}}. Springer Science \& Business Media.

\bibitem[\citeproctext]{ref-lee_training_2009}
Lee, David S. 2009. {``Training, {Wages}, and {Sample} {Selection}:
{Estimating} {Sharp} {Bounds} on {Treatment} {Effects}.''} \emph{The
Review of Economic Studies} 76 (3): 1071--1102.
\url{https://www.jstor.org/stable/40247633}.

\bibitem[\citeproctext]{ref-li_unbiased_2012}
Li, Lihong, Wei Chu, John Langford, Taesup Moon, and Xuanhui Wang. 2012.
{``An {Unbiased} {Offline} {Evaluation} of {Contextual} {Bandit}
{Algorithms} with {Generalized} {Linear} {Models}.''} In
\emph{Proceedings of the {Workshop} on {On}-Line {Trading} of
{Exploration} and {Exploitation} 2}, 19--36. JMLR Workshop; Conference
Proceedings. \url{https://proceedings.mlr.press/v26/li12a.html}.

\bibitem[\citeproctext]{ref-luedtke_statistical_2016}
Luedtke, Alexander R., and Mark J. van der Laan. 2016. {``Statistical
{Inference} for the {Mean} {Outcome} {Under} a {Possibly} {Non}-{Unique}
{Optimal} {Treatment} {Strategy}.''} \emph{The Annals of Statistics} 44
(2): 713--42. \url{https://www.jstor.org/stable/43818626}.

\bibitem[\citeproctext]{ref-manski_identification_2000}
Manski, Charles F. 2000. {``Identification Problems and Decisions Under
Ambiguity: {Empirical} Analysis of Treatment Response and Normative
Analysis of Treatment Choice.''} \emph{Journal of Econometrics} 95 (2):
415--42. \url{https://doi.org/10.1016/S0304-4076(99)00045-7}.

\bibitem[\citeproctext]{ref-manski_statistical_2004}
---------. 2004. {``Statistical {Treatment} {Rules} for {Heterogeneous}
{Populations}.''} \emph{Econometrica} 72 (4): 1221--46.
\url{https://doi.org/10.1111/j.1468-0262.2004.00530.x}.

\bibitem[\citeproctext]{ref-manski_social_2005}
---------. 2005. \emph{Social Choice with Partial Knowledge of Treatment
Response}. Econometric {Institute} Lectures. Princeton: Princeton
University Press.

\bibitem[\citeproctext]{ref-manski_identification_2007}
---------. 2007a. \emph{Identification for Prediction and Decision}.
Cambridge, Mass: Harvard University Press.

\bibitem[\citeproctext]{ref-manski_minimax-regret_2007}
---------. 2007b. {``Minimax-Regret Treatment Choice with Missing
Outcome Data.''} \emph{Journal of Econometrics}, Endogeneity,
instruments and identification, 139 (1): 105--15.
\url{https://doi.org/10.1016/j.jeconom.2006.06.006}.

\bibitem[\citeproctext]{ref-manski_treatment_2019}
---------. 2019. {``Treatment {Choice} {With} {Trial} {Data}:
{Statistical} {Decision} {Theory} {Should} {Supplant} {Hypothesis}
{Testing}.''} \emph{The American Statistician} 73 (sup1): 296--304.
\url{https://doi.org/10.1080/00031305.2018.1513377}.

\bibitem[\citeproctext]{ref-manski_identification_2024}
---------. 2024. {``{IDENTIFICATION} {AND} {STATISTICAL} {DECISION}
{THEORY}.''} \emph{Econometric Theory}, May, 1--17.
\url{https://doi.org/10.1017/s0266466624000197}.

\bibitem[\citeproctext]{ref-manski_sufficient_2016}
Manski, Charles F., and Aleksey Tetenov. 2016. {``Sufficient Trial Size
to Inform Clinical Practice.''} \emph{Proceedings of the National
Academy of Sciences} 113 (38): 10518--23.
\url{https://doi.org/10.1073/pnas.1612174113}.

\bibitem[\citeproctext]{ref-meager_understanding_2019}
Meager, Rachael. 2019. {``Understanding the {Average} {Impact} of
{Microcredit} {Expansions}: {A} {Bayesian} {Hierarchical} {Analysis} of
{Seven} {Randomized} {Experiments}.''} \emph{American Economic Journal:
Applied Economics} 11 (1): 57--91.
\url{https://doi.org/10.1257/app.20170299}.

\bibitem[\citeproctext]{ref-meager_aggregating_2022}
---------. 2022. {``Aggregating {Distributional} {Treatment} {Effects}:
{A} {Bayesian} {Hierarchical} {Analysis} of the {Microcredit}
{Literature}.''} \emph{American Economic Review} 112 (6): 1818--47.
\url{https://doi.org/10.1257/aer.20181811}.

\bibitem[\citeproctext]{ref-qian_performance_2011}
Qian, Min, and Susan A. Murphy. 2011. {``Performance {Guarantees} for
{Individualized} {Treatment} {Rules}.''} \emph{The Annals of Statistics}
39 (2): 1180--210. \url{https://www.jstor.org/stable/29783670}.

\bibitem[\citeproctext]{ref-schlag_eleven_2006}
Schlag, Karl H. 2006. {``{ELEVEN} - {Tests} {Needed} for a
{Recommendation}.''} Working \{Paper\}. European University Institute.
\url{https://cadmus.eui.eu/handle/1814/3937}.

\bibitem[\citeproctext]{ref-schochet_does_2008}
Schochet, Peter Z, John Burghardt, and Sheena McConnell. 2008. {``Does
{Job} {Corps} {Work}? {Impact} {Findings} from the {National} {Job}
{Corps} {Study}.''} \emph{American Economic Review} 98 (5): 1864--86.
\url{https://doi.org/10.1257/aer.98.5.1864}.

\bibitem[\citeproctext]{ref-semenova_generalized_2023}
Semenova, Vira. 2023. {``Generalized {Lee} {Bounds}.''} arXiv.
\url{https://doi.org/10.48550/arXiv.2008.12720}.

\bibitem[\citeproctext]{ref-stoye_minimax_2009}
Stoye, Jörg. 2009. {``Minimax Regret Treatment Choice with Finite
Samples.''} \emph{Journal of Econometrics} 151 (1): 70--81.
\url{https://doi.org/10.1016/j.jeconom.2009.02.013}.

\bibitem[\citeproctext]{ref-stoye_statistical_2011}
---------. 2011. {``Statistical Decisions Under Ambiguity.''}
\emph{Theory and Decision} 70 (2): 129--48.
\url{https://doi.org/10.1007/s11238-010-9227-2}.

\bibitem[\citeproctext]{ref-stoye_minimax_2012}
---------. 2012. {``Minimax Regret Treatment Choice with Covariates or
with Limited Validity of Experiments.''} \emph{Journal of Econometrics},
Annals {Issue} on {``{Identification} and {Decisions},''} in {Honor} of
{Chuck} {Manski}'s 60th {Birthday}, 166 (1): 138--56.
\url{https://doi.org/10.1016/j.jeconom.2011.06.012}.

\bibitem[\citeproctext]{ref-strehl_learning_2010}
Strehl, Alex, John Langford, Sham Kakade, and Lihong Li. 2010.
{``Learning from {Logged} {Implicit} {Exploration} {Data}.''} arXiv.
\url{https://doi.org/10.48550/arXiv.1003.0120}.

\bibitem[\citeproctext]{ref-swaminathan_counterfactual_2015}
Swaminathan, Adith, and Thorsten Joachims. 2015. {``Counterfactual
{Risk} {Minimization}: {Learning} from {Logged} {Bandit} {Feedback}.''}
arXiv. \url{https://doi.org/10.48550/arXiv.1502.02362}.

\bibitem[\citeproctext]{ref-tetenov_statistical_2012}
Tetenov, Aleksey. 2012. {``Statistical Treatment Choice Based on
Asymmetric Minimax Regret Criteria.''} \emph{Journal of Econometrics},
Annals {Issue} on {``{Identification} and {Decisions},''} in {Honor} of
{Chuck} {Manski}'s 60th {Birthday}, 166 (1): 157--65.
\url{https://doi.org/10.1016/j.jeconom.2011.06.013}.

\bibitem[\citeproctext]{ref-thomas_data-efficient_2016}
Thomas, Philip, and Emma Brunskill. 2016. {``Data-{Efficient}
{Off}-{Policy} {Policy} {Evaluation} for {Reinforcement} {Learning}.''}
In \emph{Proceedings of {The} 33rd {International} {Conference} on
{Machine} {Learning}}, edited by Maria Florina Balcan and Kilian Q.
Weinberger, 48:2139--48. Proceedings of {Machine} {Learning} {Research}.
New York, New York, USA: PMLR.
\url{https://proceedings.mlr.press/v48/thomasa16.html}.

\bibitem[\citeproctext]{ref-vandervaart_asymptotic_1998}
Van der Vaart, A. W. 1998. \emph{Asymptotic {Statistics}}. Cambridge
{Series} in {Statistical} and {Probabilistic} {Mathematics}. Cambridge:
Cambridge University Press.
\url{https://doi.org/10.1017/CBO9780511802256}.

\bibitem[\citeproctext]{ref-wald_contributions_1939}
Wald, Abraham. 1939. {``Contributions to the {Theory} of {Statistical}
{Estimation} and {Testing} {Hypotheses}.''} \emph{The Annals of
Mathematical Statistics} 10 (4): 299--326.
\url{https://doi.org/10.1214/aoms/1177732144}.

\bibitem[\citeproctext]{ref-wald_statistical_1945}
---------. 1945. {``Statistical {Decision} {Functions} {Which}
{Minimize} the {Maximum} {Risk}.''} \emph{The Annals of Mathematics} 46
(2): 265. \url{https://doi.org/10.2307/1969022}.

\bibitem[\citeproctext]{ref-wald_statistical_1971}
---------. 1971. \emph{Statistical Decision Functions}. 2d ed. Bronx,
N.Y: Chelsea Pub. Co.

\bibitem[\citeproctext]{ref-zambrano_social_2024}
Zambrano, Eduardo. 2024. {``Social {Preferences} {Under} {Uncertainty}
and {Ambiguity}.''}

\bibitem[\citeproctext]{ref-zhang_estimating_2012}
Zhang, Baqun, Anastasios A. Tsiatis, Marie Davidian, Min Zhang, and Eric
Laber. 2012. {``Estimating Optimal Treatment Regimes from a
Classification Perspective.''} \emph{Stat} 1 (1): 103--14.
\url{https://doi.org/10.1002/sta.411}.

\bibitem[\citeproctext]{ref-zhao_estimating_2012}
Zhao, Yingqi, Donglin Zeng, A. John Rush, and Michael R. Kosorok. 2012.
{``Estimating {Individualized} {Treatment} {Rules} {Using} {Outcome}
{Weighted} {Learning}.''} \emph{Journal of the American Statistical
Association} 107 (499): 1106--18.
\url{https://www.jstor.org/stable/23427417}.

\end{CSLReferences}
\end{document}